%% file: main.tex
\documentclass[sigconf]{acmart}
\usepackage{booktabs}
\usepackage{multirow}
\usepackage{graphicx}
\usepackage{natbib}
\usepackage{subcaption}
\usepackage{amsthm}

\newtheorem{property}{Property}
\usepackage{pbox}
\usepackage{pifont}
\usepackage{cleveref}
\usepackage[normalem]{ulem}
\useunder{\uline}{\ul}{}
\AtBeginDocument{%
  }

\copyrightyear{2025}
\acmYear{2025}
\setcopyright{cc}
\setcctype{CC-BY}
\acmConference[WSDM '25]{Proceedings of the Eighteenth ACM International Conference on Web Search and Data Mining}{March 10--14, 2025}{Hannover, Germany}
\acmBooktitle{Proceedings of the Eighteenth ACM International Conference on Web Search and Data Mining (WSDM '25), March 10--14, 2025, Hannover, Germany}\acmDOI{10.1145/3701551.3703587}
\acmISBN{979-8-4007-1329-3/25/03}

\makeatletter
\gdef\@copyrightpermission{
\begin{minipage}{0.3\columnwidth} \href{https://creattvecommons.org/licenses/by/4.0/}{\includegraphics[width=0.90\textwidth]{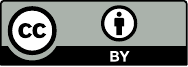}}
\end{minipage}\hfill \begin{minipage}{0.7\columnwidth}
\href{https://creattvecommons.org/licenses/by/4.0/}{This work is licensed under a Creative Commons Attribution International 4.0 License.}
\end{minipage}
\vspace{5pt}
}
\makeatother

\begin{document}

\title{Your Causal Self-Attentive Recommender Hosts a Lonely Neighborhood}

\author{Yueqi Wang}
\authornote{Both authors contributed equally to this research.}
\affiliation{%
  \institution{University of California, Berkeley}
  \city{Berkeley}
  \country{USA}
}
\email{yueqi@berkeley.edu}

\author{Zhankui He}
\authornotemark[1]
\affiliation{%
  \institution{University of California, San Diego}
  \city{San Diego}
  \country{USA}
}
\email{zhh004@eng.ucsd.edu}

\author{Zhenrui Yue}
\affiliation{%
  \institution{University of Illinois Urbana-Champaign}
  \city{Champaign}
  \country{USA}
}
\email{zhenrui3@illinois.edu}

\author{Julian McAuley}
\affiliation{%
  \institution{University of California, San Diego}
  \city{San Diego}
  \country{USA}
}
\email{jmcauley@eng.ucsd.edu}

\author{Dong Wang}
\affiliation{%
  \institution{University of Illinois Urbana-Champaign}
  \city{Champaign}
  \country{USA}
}
\email{dwang24@illinois.edu}

\renewcommand{\shortauthors}{Yueqi Wang, Zhankui He, Zhenrui Yue, Julian McAuley, \& Dong Wang}

\begin{abstract}
In the context of sequential recommendation, a pivotal issue pertains to the comparative analysis between bi-directional/auto-enco-\\ding (AE) and uni-directional/auto-regressive (AR) attention mechanisms, where the conclusions regarding architectural and performance superiority remain inconclusive. Previous efforts in such comparisons primarily involve summarizing existing works to identify a consensus or conducting ablation studies on peripheral modeling techniques, such as choices of loss functions. However, far fewer efforts have been made in (1) theoretical and (2) extensive empirical analysis of the self-attention module, the very pivotal structure on which performance and designing insights should be anchored. In this work, we first provide a comprehensive theoretical analysis of AE/AR attention matrix in the aspect of (1) sparse local inductive bias, a.k.a neighborhood effects, and (2) low rank approximation. Analytical metrics reveal that the AR attention exhibits sparse neighborhood effects suitable for generally sparse recommendation scenarios. Secondly, to support our theoretical analysis, we conduct extensive empirical experiments on comparing AE/AR attention on five popular benchmarks with AR performing better overall. Empirical results reported are based on our experimental pipeline named \underline{\textbf{Mo}}dularized \underline{\textbf{D}}esign Space for \underline{\textbf{S}}elf-\underline{\textbf{A}}ttentive \underline{\textbf{R}}ecommender (\textbf{ModSAR}), supporting adaptive hyperparameter tuning, modularized design space and Huggingface plug-ins. We invite the recommendation community to utilize/contribute to ModSAR to (1) conduct more module/model-level examining beyond AE/AR comparison and (2) accelerate state-of-the-art model design. Lastly, we shed light on future design choices for performant self-attentive recommenders. We make our pipeline implementation and data available at https://github.com/yueqirex/SAR-Check.
\end{abstract}




\maketitle

\input{1_intro}
\input{5_preliminary_and_setup}
\input{2_Sparsity_Measure}
\input{3_Low_rank_approx}
\input{4_Large_scale_empirical_results}

\section{Acknowledgments}
This research is supported in part by the National Science Foundation under Grant No. CNS-2427070,  IIS-2331069, IIS-2130263, CNS-2131622. The views and conclusions contained in this document are those of the authors and should not be interpreted as representing the official policies, either expressed or implied, of the U.S. Government. The U.S. Government is authorized to reproduce and distribute reprints for Government purposes notwithstanding any copyright notation here on.

\bibliographystyle{ACM-Reference-Format}
\balance
\bibliography{sample-base}

\end{document}

%% file: 1_intro.tex
\section{Introduction}
\label{sec:introduction}

``\textit{Auto-Encoding} (AE) or \textit{Auto-Regression} (AR)?'': The debate over model architecture has existed in the context of sequence modeling tasks since the introduction of transformer models~\cite{vaswani2017attention, devlin2018bert}. For language tasks, this discourse has driven the prosperity of BERT-like and GPT-like models, and the recent advances in Large Language Models (LLMs) show a milestone for AR models~\cite{devlin2018bert, radford2018improving, brown2020language, touvron2023llama}. In the field of sequential recommendation that aims to predict users' next action based on their historical action sequences, both AE and AR self-attentive models have achieved state-of-the-art performance~\cite{kang2018self, sun2019bert4rec, wu2020sse}. Nevertheless, in the context of sequential recommendations, the ``\emph{AE} or \emph{AR} attention'' question still remains challenging to answer, especially regarding theoretical analysis and interpretation of attention behaviors.

AR-based and AE-based self-attentive models are widely used in sequential recommendation tasks~\cite{liu2021noninvasive, petrov2021attention, fan2021lighter}. AR-based recommenders, represented by SASRec~\cite{kang2018self}, SSE-PT~\cite{wu2020sse}, are commonly compared with AE-based recommenders, represented by BERT4Rec \cite{sun2019bert4rec}, LOCKER~\cite{he2021locker} in a wide range of recommendation datasets. However, those comparisons between AR-based and AE-based recommenders always lead to discrepancies in conclusions from a number of papers~\cite{petrov2022systematic}. Specifically, while AE models claim superiority of bi-directional attention over AR models in~\cite{ma2020disentangled, zhou2021contrastive, liu2021augmenting, li2021intention, cho2020meantime, zhang2020match4rec, tong2021pattern, yue2021black, wu2021seq2bubbles, sun2022sequential}, AR models show better performance in~\cite{zhou2020s3, li2021lightweight, zhang2021behavioral, qiu2022contrastive}. Approaching this issue, \cite{petrov2022systematic} focuses on replicating the original results of the AE-based recommender BERT4Rec, rectifying several implementation details to obtain closer results to the reported. Afterwards, \cite{klenitskiy2023turning} shows that the AR-based recommender, SASRec, outperforms the AE-based BERT4Rec when using the same cross-entropy loss function of the latter.

While existing works of comparing AE~/~AR are mainly focused on statistically summarizing past works and studying peripheral modeling/evaluation techniques, we argue that a more comprehensive conclusion should be pivoting around the behavior of AE~/~AR self-attention which serves as the key component of self-attentive sequential recommenders. To address this issue, below we formulate three specific research questions and share a summary of our findings and insights:

\textbf{Q1: What are proper theoretical measurements that reflect behaviors of self-attention?}
Theoretical measurements are vital regarding interpreting attention behaviors. However, constructing proper measurements remains challenging since improper ones often lead to deviated patterns and misunderstandings. To solve this issue, we carefully choose two measurements of attention matrices regarding (1) sparsity (2) rank-$k$ approximation. For both aspects, we provide mathematical properties or accuracy upper bounds showing why they are decent measurements.

\textbf{Q2: What do the measurements and visualizations tell us about traits of AE~/~AR attention?}
It remains mysteriously exciting how self-attention behaves in sequential recommendation. We provide visualization and interpretations to pair with our theoretical measurements. We share our findings in three aspects, namely (1) attention visualization, (2) singular value visualization and (3) performance visualization.

\textbf{Q3: Does the theoretical advantage of AE~/~AR models transform into empirical performance?}
In parallel with theoretical measurements, we are also curious whether AE~/~AR model exhibit performance superiority over the other. To include more model variants for inclusive conclusions, we provide performance reports of AE~/~AR in a large-scale experiment of (1) vanilla attention comparison, (2) variant attention comparison in common design space for sequential recommendation, (3) variant attention comparison in the broader Huggingface community including more advanced AE~/~AR models.

We introduce preliminaries and overall setups in \Cref{sec: prelim and overall setup}. We provide three findings corresponding to the above proposed research questions in \Cref{sec: sparsity measure}, \Cref{sec: low rank approx} and \Cref{sec: emprical exp}. Specifically, within each finding, we provide a theoretical analysis followed by interpretation and visualization. Empirical performance report is discussed in \Cref{sec: emprical exp}. Connection between theoretical and empirical analysis is discussed in \Cref{sec: connect empirical and theory}. We finally discuss conclusions in \Cref{sec: conclusion}. In addition to theoretical and empirical results, we make our code and experimental scripts publicly available to support future research with a link attached to the end of the abstract.

%% file: 5_preliminary_and_setup.tex
\section{Preliminaries and Overall Setup}
\label{sec: prelim and overall setup}
\subsection{Preliminaries}
\noindent\textbf{Self-attention}: The self-attention mechanism is used to process item correlations in a sequence where $X \in \mathbb{R}^{L \times H}$ is $L$-length sequence of $H$-dimensional token embeddings. Learnable weights $W_Q, W_K, W_V$ are used to transform X into $Q,K,V$ and compute attention correlation exerted to $V$:
\begin{equation}
    \text{Attn(X)} = softmax(\frac{(XW_Q)(XW_K)^T}{\sqrt{d_k}}) \times (XW_V)
\end{equation}

\noindent\textbf{Sequential recommendation}: The sequential recommendation task is formulated as follows. Consider a user set $U=\{u1, u2,...,u_{|U|}\}$, an item set $V=\{v1, v2,...,v_{|V|}\}$ and a user $u_i$'s interacted item sequence in chronological order $S_u=[v_1^{(u)}, v_2^{(u)},...,v_n^{(u)}]$ with $n$ being the sequence length, the sequential recommendation task is to predict the next item $v_{n+1}^{(u)}$ that $u$ will interact with. Formally, it could written as the probability of each item $v \in V$ being the the next interacted item given $S_u$:
\[ p(v_{n+1}^{(u)}=v|S_u) \]

\noindent\textbf{Matrix sparsity measurement}: Given a matrix/vector $\boldsymbol{X} \in \mathbb{R}^{m \times n}$ or $\boldsymbol{x} \in \mathbb{R}^n$, a sparsity measure $S: \mathbb{R}^n \to \mathbb{R}, \boldsymbol{x} \mapsto S(\boldsymbol{x})$ outputs a \textit{scalar value} representing the sparsity of $\boldsymbol{x}$.

\noindent\textbf{SVD-based rank-$k$ matrix approximation}: Given a matrix $\boldsymbol{X} \in \mathbb{R}^{m \times n}$, a SVD-based matrix approximation is using largest $k$ singular values and vectors to form a rank-$k$ approximation $\boldsymbol{X}_k=U_k \Sigma_k V_k$, where $U \in \mathbb{R}^{n \times k}, \Sigma \in \mathbb{R}^{k \times k}, V \in \mathbb{R}^{n \times k}$.

\subsection{Overall Setup}
\subsubsection{Datasets}
\label{ssec: datasets}
We select five popular datasets for (1) e-commerce recommendation (Beauty, Sports, Video), (2) local business recommendation (Yelp) and (3) movie recommendation (ML-1M). Dataset selection is based on a survey of 48 most-cited papers of sequential recommendation and we use ones with most frequent appearances~\cite{kang2018self, sun2019bert4rec, zhou2022filter, wu2020sse}. Specifically, Beauty, ML-1M, Yelp, Sports, Video appears 20 (41.67\%), 18 (37.50\%), 10 (20.83\%), 9 (18.75\%), 9 (18.75\%) times of the 48 surveyed papers:
\begin{table}[t]
    \footnotesize
    \caption{Dataset statistics after preprocessing.}
    \centering
    \begin{tabular}{lrrccc}
    \toprule
    \textbf{Datasets} & \textbf{Users} & \textbf{Items} & \textbf{Interact.} & \textbf{Length} & \textbf{Density} \\ \midrule
    \textbf{Beauty}   & 52,204         & 57,289         & 395K               & 7.6               & 1e-4             \\
    \textbf{Sports}   & 83,970         & 83,728         & 589K               & 7.0               & 8e-5             \\
    \textbf{Video}    & 31,013         & 23,715         & 287K               & 9.3               & 4e-4             \\
    \textbf{Yelp}   & 31,371         & 36,616         & 300k               & 9.6               & 3e-4             \\
    \textbf{ML-1M}    & 6,040          & 3,416          & 1M                 & 165.5             & 5e-2             \\
    \bottomrule
    \end{tabular}
    \label{tab:dataset}
    \vspace{-15pt}
\end{table}
For preprocessing, we follow common practice in~\cite{kang2018self, he2021locker, he2017neural, tang2018personalized, rendle2010factorizing} that remove users and items with fewer than 5 interactions. We do not use k-core filtering to keep the natural sparsity of datasets~\cite{kang2018self, he2021locker}. We adopt a max sequence length of 200 for ML-1M and 50 for others. We use the last item in the sequence as the test set, the second to last item for validation, and the rest of the items for training. Data statistics after preprocessing are in Table~\ref{tab:dataset}.

\subsubsection{Baseline Models}
\noindent\textbf{Vanilla AE/AR attention models} used in \Cref{sec: sparsity measure} and \Cref{sec: low rank approx} for theoretical analysis and \Cref{sec: exp: vanilla attn} for performance comparison: \textbf{(1) Vanilla-AE}: the vanilla auto-encoding-based attention model. We adopt BERT4Rec~\cite{sun2019bert4rec}, which is a classic pure bi-directional attention model for recommendation. \textbf{(2) Vanilla-AR}, the vanilla auto-aggressive attention model. We adopt SASRec~\cite{kang2018self} which is a pure unidirectional self-attentive recommender with causal mask meaning no look to future items. \textbf{(3) Local/Local-attn}: we adopt the local attention mechanism in ~\cite{he2021locker} that applies a fixed-size window constraint to vanilla bi-directional attention.

\textbf{Variant AE/AR attention models} as common model design choices used in \Cref{sec: exp common design space} for more inclusively comparing AE~/~AR's performance superiority (Local-attn models are also included and are already introduced above): \textbf{(1) SSE-PT}: A self-attentive recommender that introduces user modeling in addition to item embedding with random dropout. \textbf{(2) Loss-X}: The vanilla-AE/AR model using loss-X for optimization.

\textbf{Variant AE/AR attention models} as in broader Huggingface AI community used in \Cref{sec: exp huggingface} for more inclusively comparing AE~/~AR's performance superiority. Local-attn models are also included and are already introduced above: \textbf{(1) LLaMA}: A variant transformer architecture with optimized layer configuration and normalization techniques. \textbf{(2) ALBERT}: An optimized BERT architecture to reduce model size and computational requirements while maintaining high performance through parameter sharing and factorized embedding. \textbf{(3) Trm.-XL}: An unidirectional transformer with segment-level recurrence and a memory mechanism to handle long-range dependencies and context more effectively than standard Transformers.

\subsubsection{Implementation Details}
\label{ssec: eval and reprod std}
\noindent\textbf{Training and evaluation} The evaluation metrics for all experiments are Recall@$k$ and NDCG@$k$ ($k \in [5,10,20]$), which are the most commonly used metrics in related works. We adopt Cross-Entropy (CE) loss when not explicitly mentioned otherwise and set max epochs to 1000 and early stop when Recall@10 does not improve for 20 epochs. Each one of all models is trained on a single NVIDIA-3090 GPU (24G). The training time of each model ranges from 40 minutes to 3 hours depending on the dataset.

\noindent\textbf{Hyper-parameter tuning} For hyper-parameter tuning, we apply grid search for models in \Cref{sec: exp: vanilla attn} and ray-tune managed adaptive search for ones in \Cref{sec: exp common design space} and \Cref{sec: exp huggingface} for more inclusive and accurate results comparison given limited number of machines. We search the hidden size of BERT4Rec in [32, 64, 128, 256, 512]. We search the hidden activation dropout rate and attention dropout rate from 0.1 to 0.9 on a 0.1 stride. We also search the number of attention layers and attention heads of [1,2,4]. Lastly, we search the masking probability of bidirectional attention models among [0.2, 0.4, 0.6, 0.8]. Due to the page limit, we do not enumerate through the tuning of all hyperparameters but rather the above important ones. Refer to our shared code base for more details.

%% file: 2_Sparsity_Measure.tex
\section{Finding 1: Sparse Local Inductive Bias of AR Attention}
\label{sec: sparsity measure}
Finding 1: Sparse local inductive bias of AR attention is the key performance contributor

In this section, we introduce how to effectively measure the distribution of the attention matrix in AE and AR models to study models' behavior. One way to measure such distribution is (i) \textit{sparsity measurement} with (ii) \textit{sparsity locating} to quantitatively surface how much the attention matrix biases towards certain regions and where. In theory, sparsity measurement poses significant challenges. Intuitively speaking, a matrix with more zeros tends to be sparser. However, attention matrices are usually positive due to softmax re-scale, making it vital to measure other aspects of sparsity such as uneven distribution of attention scores which represent an emphasis on certain areas, etc.\cite{brown2020language}. Inspired by \cite{wang2010information}, we adapt the sparsity measurement theory to analyze the attention matrix in self-attention-based recommendation models. To provide a comprehensive measurement with ideal traits, we first derive an optimal sparsity measurement and its properties. Secondly, we analyze the sparsity analysis with attention visualization.

\subsection{Smooth Sparsity Measurement of Attention Matrix}
For the attention matrix, we first review (1) why traditional sparsity measurement falls short and (2) the mathematical derivation of a more scientifically adopted metric proposed by~\cite{kexuefm-9595}. Sparsity traditionally~\cite{hurley2009comparing} refers to the proportion of zero elements relative to the total. However, for attention matrices, the concept differs: sparsity signifies that most entries are negligible or near zero, with only a minority exhibiting substantial values. Quantifying sparsity in attention matrices necessitates a clear definition of what constitutes a "significant" value. A practical method involves computing the average absolute value of all elements, normalized by the maximum absolute value:
\begin{equation}
\label{eq: naive measure}
    S(\boldsymbol{x}) = \frac{(|x_1|+|x_2|+\ldots+|x_n|)/n}{\max({|x_1|,|x_2|,\ldots,|x_n|})}
\end{equation}

The introduction of the max operation in \Cref{eq: naive measure} compromises smoothness, rendering the measure highly susceptible to outliers. Leveraging approximation theory, the \textit{LogSumExp (LSE)} function, defined in \Cref{eq: LSE approx}, offers a smooth alternative to the maximum function. Within the attention matrix $X$, this inherent smoothness effectively attenuates the influence of extreme values~\cite{kexuefm-9595}.
\begin{equation}
\label{eq: LSE approx}
\max(|x_1|,|x_2|,\ldots,|x_n|)\approx \frac{1}{n}\log\sum_{i=1}^n e^{|x_i|}
\end{equation}

Nevertheless, the $e^{|x_i|}$ operation in the \textit{LSE} function introduces amplification effects for large values within the matrix. To address this, $e^{|x_i|}$ in \Cref{eq: LSE approx} is replaced by $|x_i|^p$, where $p$ is typically set to 2 and the averaging factor $1/n$ is adjusted to $1/p$ in its power. Thus far, we have

\begin{equation}
\label{eq: 2nd approx}
    \max(|x_1|,|x_2|,\ldots,|x_n|)\approx \left(\sum_{i=1}^n |x_i|^p\right)^{1/p}\triangleq L_p(\boldsymbol{x})
\end{equation}

Observe that $(|x_1|+|x_2|+\ldots+|x_n|)/n$ corresponds to the $L_1$-norm of $\mathbf{x}$. Substituting the $L_1$-norm and \Cref{eq: 2nd approx} into \Cref{eq: naive measure} yields \Cref{eq:s1p}. Empirically, $p$ usually takes the values of 2 and omitting the scaling factor of $n$ given fixed size vector/matrix comparison. Under these assumptions, the resulting sparsity measure~\cite{kexuefm-9595} exhibits several key properties, which we will explore in greater detail.

\begin{equation}S(\boldsymbol{x})=\frac{L_1(\boldsymbol{x})/n}{L_p(\boldsymbol{x})}\label{eq:s1p}\end{equation}

The sparsity measurement theory \cite{hurley2009comparing, kexuefm-9595} outlines six essential properties for effectively capturing the nuanced sparsity of a vector $\boldsymbol{x}$. Among these, two properties resonate closely with the established framework of the attention matrix. The attention sparsity measure employed in this study is rigorously proven to satisfy these two properties. These are discussed as follows:

Specifically, for a non-negative vector $\boldsymbol{x}$, an ideal sparsity measure $S: \mathbb{R}^n \to \mathbb{R}, \boldsymbol{x} \mapsto S(\boldsymbol{x})$ must adhere to two fundamental conditions:
\begin{property}
\label{property 1}
Given two elements $x_i > x_j$ of real-valued vector $\boldsymbol{x} \in \mathbb{R}^n$ and a positive real number $0 < \alpha < \frac{x_i - x_j}{2}$, the measurement $S$ of $\boldsymbol{x}$ should exhibit $S([\ldots,x_i - \alpha,\ldots,x_j + \alpha,\ldots]) > S(\boldsymbol{x})$.
\end{property}

\begin{property}
\label{property 2}
Given $\forall i \in {1,2,\ldots,n}, \exists \beta>0$, such that $\forall \alpha>0$, it is true that $S([\ldots,x_{i-1},x_i + \beta_i + \alpha,x_{i 
+1},\ldots]) < S([\ldots,x_{i-1},x_i + \beta_i,x_{i+1},])$
\end{property}

Property \ref{property 1} implies that, given a fixed total sum of the vector, greater unevenness in the distribution of $\boldsymbol{x}$ leads to increased sparsity. This mirrors the regional focus of the attention matrix on more informative tokens or items. Property \ref{property 2} asserts that when specific elements in $\boldsymbol{x}$ reach a sufficient magnitude, the overall sparsity is predominantly governed by those larger values. This is desirable, as it reflects the significant influence of larger attention values in generating high-quality recommendations.

\begin{table}[t]
\caption{L1/L2 sparsity computation, which measures the unevenness of attention matrix other than counting zero/near-zero elements. Smaller values denote higher sparsity meaning more attention is focused on a few important elements.}
\centering
\resizebox{\linewidth}{!}{%
\begin{tabular}{@{}lcccccc@{}}
\toprule
\textbf{Model/Dataset}                   & \textbf{Beauty} & \textbf{Sports} & \textbf{Video} & \textbf{Yelp} & \textbf{ML-1M} & \textbf{Avg.} \\ \midrule
\multicolumn{1}{l|}{\textbf{Vanilla-AE}} & 0.018           & 0.013           & 0.021          & 0.019         & 0.038          & 0.022         \\
\multicolumn{1}{l|}{\textbf{Local-AE}}   & 0.014           & 0.009           & 0.017          & 0.015         & 0.033          & 0.018         \\
\multicolumn{1}{l|}{\textbf{Vanilla-AR}} & 0.013           & 0.007           & 0.015          & 0.013         & 0.030          & 0.016         \\ \bottomrule
\end{tabular}%
}
\label{tab: l1/l2 sparsity}
\end{table}

\begin{figure}[t]
  \centering
  \includegraphics[width=\linewidth]{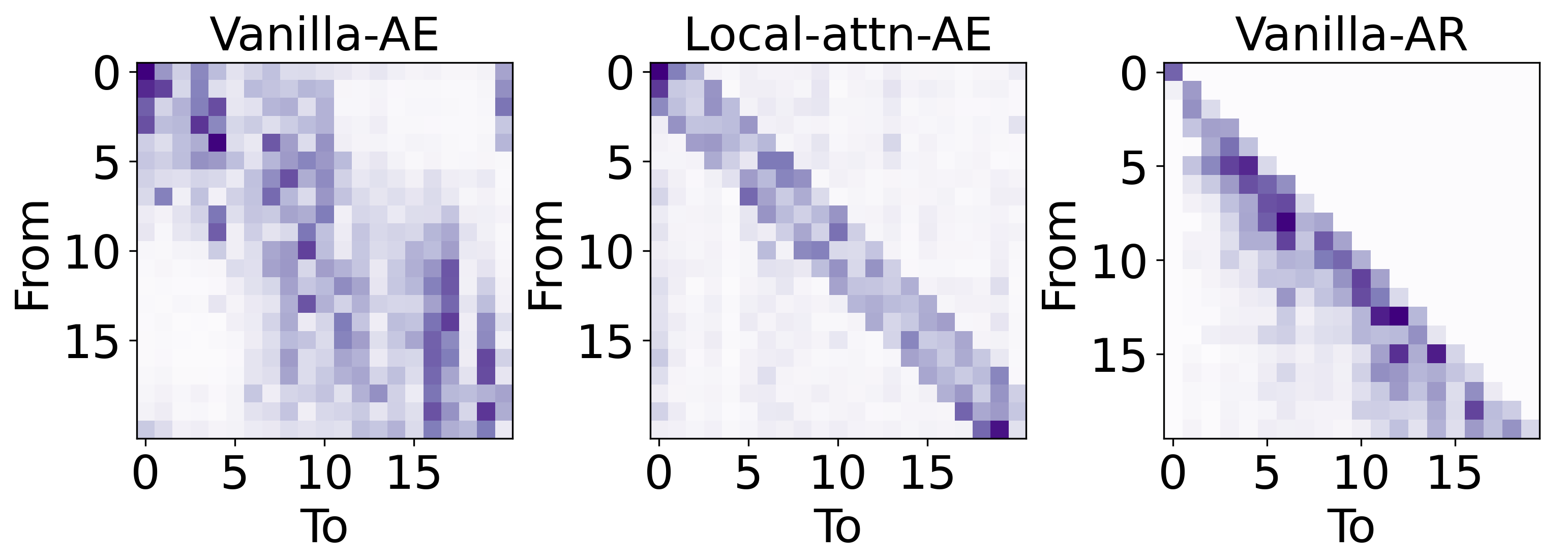}
  \caption{Attention visualizations. local effect grows from left to right.}
  \label{fig:attn_visual_1row}
  \vspace{-10pt}
\end{figure}

\subsection{Sparsity Analysis, Visualization and Interpretability}
We follow the L1/L2 sparsity computation introduced in the above section and report the results in \Cref{tab: l1/l2 sparsity}. Smaller values denote higher sparsity. Recall that our L1/L2 sparsity metrics emphasize on unevenness, a.k.a relative large/small values than others in the attention matrix, rather than directly counting zeros/near-zero elements in the matrix. \Cref{tab: l1/l2 sparsity} shows that the Vanilla-AR attention matrix possesses the highest sparsity which represents that large attention values are assigned to a few important tokens informative for generating recommendations. In contrast, Vanilla-AE models show the lowest sparsity meaning that attention scores are more evenly distributed across different history items. Intuitively, more specialized attention is required for aligning with the local effects in sequential recommendation, especially for sparse datasets with a short average item-purchase sequence for a user \cite{singh2020scalability}.

To further assure what the sparse AR attention is modeling compared with its AE counterparts, we further visualize the attention matrix for Vanilla-AE, Local-AE and Vanilla-AR attention in \Cref{fig:attn_visual_1row}. \Cref{fig:attn_visual_1row} shows that vanilla-AR model is indeed attending heavily to the local items of the predicted item. Overall, the Vanilla-AR models offer a "\textbf{Lonely Neighborhood}" where distant items are rarely attended to. This local effect is even higher than the deliberately constrained Local-AE model with intensive dark cells matching the diagonal (\Cref{fig:attn_visual_1row} rightest sub-figure). We call this \textit{sparse local-inductive bias} since it is a sparsely distributed matrix learned through auto-regressive fashion training where the model always predicts the next token every time. So far, we have associated the uneven attention distribution/sparse attention of Vanilla-AR to the local neighborhood effects. While this aligns with the findings in past literature of locality in recommendation benchmarks \cite{singh2020scalability, he2021locker}, we further conduct empirical experiments in \Cref{sec: emprical exp} on local attention to show performance indeed increases with higher local inductive bias, which is Vanilla-AE < Local-AE < Vanilla-AR.

%% file: 3_Low_rank_approx.tex
\section{Finding 2: Richer Data Dynamics are Stored in AR Attention} 
\label{sec: low rank approx}
In this section, we focus on another aspect of attention matrix via its rank analysis. The low-rank approximation of a matrix shows the latent representation ability of the given matrix. We first introduce how to derive a proper $k$-rank approximation and then analyze approximation results with visualizations.

\subsection{Low Rank Approximation of Attention Matrix}
In this section, we analyze the attention matrix in the aspect of matrix rank. In matrix theory, matrix rank conveys significant implications of information capacity and variance a matrix could represent \cite{gao2018low, candes2012exact}. In particular, a higher rank approximation demanded by a given matrix $A$ implies more information dynamics $A$ could carry, and a lower rank approximation required implies more redundancies of $A$ thus less information variance. By "demanded" we mean the lowest rank approximation required to retain the original matrix pattern. One effective way to conduct low-rank approximation is through Singular Value Decomposition (SVD). In particular, SVD decomposes a given matrix $A$ into singular values and vectors, the subset of which could form a low-rank approximation of the original matrix.

Formally for an attention matrix $A \in \mathbb{R}^{n \times n}$ which is usually real-valued and square, the SVD-based low-rank approximation of an original matrix  operates by first conducting the decomposition:
\begin{equation}
    A = U \Sigma V
\end{equation}
where $U \in \mathbb{R}^{n \times n}$ is an orthogonal matrix satisfying $U^TU=UU^T=I$ and $I$ is the identity matrix; $\Sigma \in \mathbb{R}^{n \times n}$ is a diagonal matrix containing all the singular values of $A$; $V \in \mathbb{R}^{n \times n}$ is also an orthogonal matrix.

Secondly, we take the first $k$ singular vectors of $U$ and $V$, the corresponding $k$ singular values in $\Sigma$, and compute the low-rank approximation $A_k$ of $A$ as $A_k=U_k \Sigma_k V_k$, where $U \in \mathbb{R}^{n \times k}, \Sigma \in \mathbb{R}^{k \times k}, V \in \mathbb{R}^{n \times k}$.

Thirdly, it is challenging to determine $k$ that produces an accurate and robust approximation of $A$. A rank $k$ too high would make trivial our goal to detect the information variance in the original matrix $A$ and a rank too low would completely destroy the patterns in $A$. Here, we adopt a hard-threshing measurement that determines k that yields a robust approximation of $A$ that is (i) robust to a certain noise level and (ii) accurately represents the original signal in $A$. Here we utilize the concept of hard-thresholding in matrix approximation theory \cite{cai2010singular}, where a noise-based threshold $\tau$ is used to filter all the $k$s above that threshold. The filtered $k$s are used for approximation using SVD matrix reconstruction. We show later this approximation accuracy is upper-bounded by a common selection of noise level $\tau$. In another word, the hard-thresholding method provides a noise-thresholding rank-$k$ approximation of the original matrix that is noise-robust and accurate. This is better than methods such as elbow-method without accuracy upper bound of the approximation. Concretely, we first construct a noise-perturbed version of $A$ as

\begin{equation}
    Y=A+E
\end{equation}

We then compute the threshold $\tau$ as
\begin{equation}
    \tau = \sqrt{2(\sigma^2 + \|\Sigma^{(Y)}\|_F^2/n^2)\log(2n)}
\end{equation}
where $\sigma$ is the variance of the noise; $\|\Sigma^{(Y)}\|_F$ is the Frobenius norm of the singular vector $\Sigma^{(Y)}$ of the noise matrix $Y$. The Frobenius is intuitively the matrix version $L_2$-norm. It is computed by $\|\Sigma^{(Y)}\|_F=\sqrt{\sum_{i=1}^m \sum_{j=1}^n |\Sigma^{(Y)}_{ij}|}$ where $m,n$ are the dimension of $Y$ and here in our case $m=n$. The reason noise $\tau$ is designed this way is that the final approximation of $A$ would satisfy an accuracy upper bound which we will show later.

After $\tau$ is derived, we count the number of singular values of $Y$ stored in $\Sigma^{(Y)}$ that are greater than $\tau$:
\begin{equation}
    k = \sum_{i=1}^n \mathbf{1}\{s_i > \tau\}, s_i \in \{\Sigma^{(Y)}_{ij}\}
\end{equation}
where $\mathbf{1}\{\cdot\}$ is the indicator function that returns 1 when the inside statement is true.

Finally, we use the $k$ selected by the noise-threshold $\tau$ to compute the approximation of the original matrix $A$ by

\begin{equation}
    A_k=U_k \Sigma_k V_k
\end{equation}
which satisfies the following upper bound:
\begin{equation}
    \| A - A_k \|_F^2 \leq \text{min}_r \left(\| A - A_r \|_F^2 + r\sigma^2 \right)
\end{equation}
which says that the squared error of the rank=$k$ approximation of $A$ is upper-bounded by the smallest-rank approximation error (which is usually large) plus squared noise variance $\sigma$.

\subsection{Approximation Results, Visualization and Interpretability}

\begin{table}[t]
\caption{Rank $k$ determined by noise-based hard thresholding method. Such rank-$k$ SVD approximations are guaranteed robust to noise and accurate by a closed-form upper bound.}
\centering
\resizebox{\linewidth}{!}{%
\begin{tabular}{@{}lrrrrrr@{}}
\toprule
\textbf{Model/Dataset}                   & \textbf{Beauty} & \textbf{Sports} & \textbf{Video} & \textbf{Yelp} & \textbf{ML-1M} & \textbf{Avg.} \\ \midrule
\multicolumn{1}{l|}{\textbf{Vanilla-AE}} & 5               & 5               & 6              & 6             & 5              & 5.4           \\
\multicolumn{1}{l|}{\textbf{Local-AE}}   & 9               & 11              & 10             & 11            & 12             & 10.6          \\
\multicolumn{1}{l|}{\textbf{Vanilla-AR}} & 8               & 9               & 9              & 8             & 9              & 8.6           \\ \bottomrule
\end{tabular}%
}
\label{tab: rank compute}
\end{table}
\input{figures/attn_visual}
\input{figures/svd}
We follow the noise-based threshold method introduced in the above section to determine the rank-$k$ and report the results in \Cref{tab: rank compute}. Recall that the noise-based threshold method ensures that the minimum $k$ is found to retain the intrinsic signal (accurate) and is robust to random noise. We find that the Vanilla-AE model requires the smallest $k$ for a meaningful SVD low-rank approximation with an average of $k=5.4$ over all datasets. In contrast, overall higher ranks are required to properly approximate Local-AE and Vanilla-AR with an average $k=10.6$ and $k=8.6$. Higher rank approximation in SVD implies more variant data dynamics stored in the matrix \cite{gao2018low}. This is true for the Local-AE and Vanilla-AR model with potentially better representation ability compared with the Vanilla-AE model that requires a lower rank approximation.

To further surface the low-rank approximation effects. We further provide a visualization in \Cref{fig:attn_visual} the approximation results of the ML-1M dataset (due to page limit) using $k=5$, which is the smallest $k$ required for well-estimating the Vanilla-AE attention. \Cref{fig:attn_visual} shows that Vanilla-AE still retains its pattern, while the Local-AE and Vanilla-AR already lost their patterns of sparse local inductive bias introduced in \Cref{sec: sparsity measure}. This is expected since the latter two require higher-rank approximations derived from our \textit{noise-based hard thresholding} method.

In addition, we sort the singular values in descending order and visualize their decreasing behavior in \Cref{fig:singular_value_show}. Due to the page limit, we only show the visualization of ML-1M and Beauty datasets here. \Cref{fig:singular_value_show} shows that for both attention layers, the Vanilla-AR exhibits a slower decay rate of its singular values compared with Vanilla-AE. According to more matrix approximation theory \cite{gao2018low, candes2012exact}, Large singular values are crucial in carrying intrinsic signals stored in the original matrix. Thus, overall more large singular values in AR attention implies stronger representation ability of user-item interaction signals.

%% file: figures/attn_visual.tex
\begin{figure}[t]
  \centering
  \includegraphics[width=\linewidth]{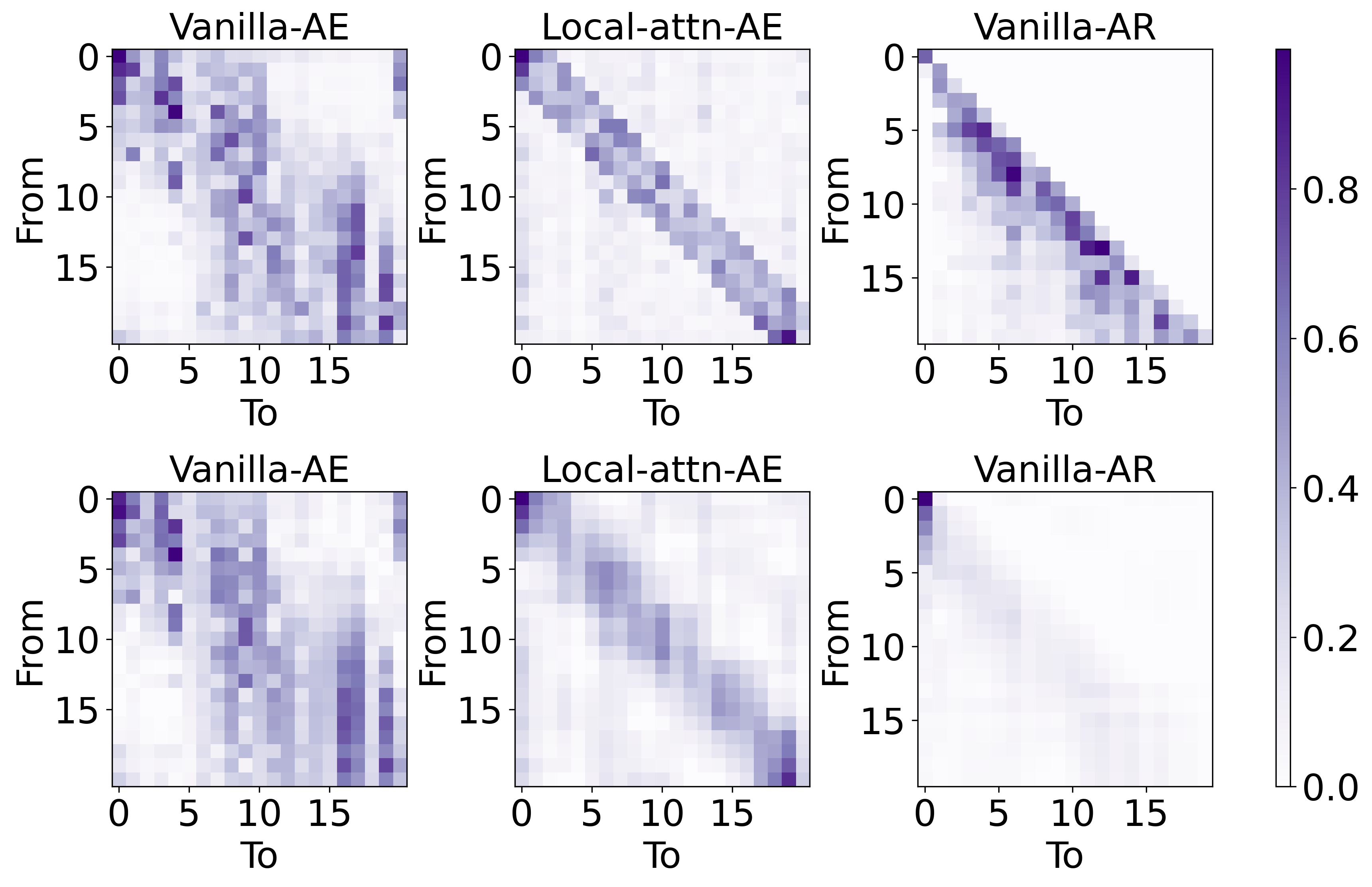}
  \caption{Attention visualizations. First row is original matrix; Second row is corresponding low-rank approximations using top-5 largest singular values. Vanilla-AE (BERT4Rec-like here) attention has a clear-pattern rank-5 approximation while local-attn and AR losses their patterns and need a higher-rank approximation.}
  \label{fig:attn_visual}
\end{figure}

%% file: figures/svd.tex
\begin{figure}[t]
    \centering
    \begin{subfigure}[]{0.49\linewidth}
        \centering
        \includegraphics[width=\textwidth]{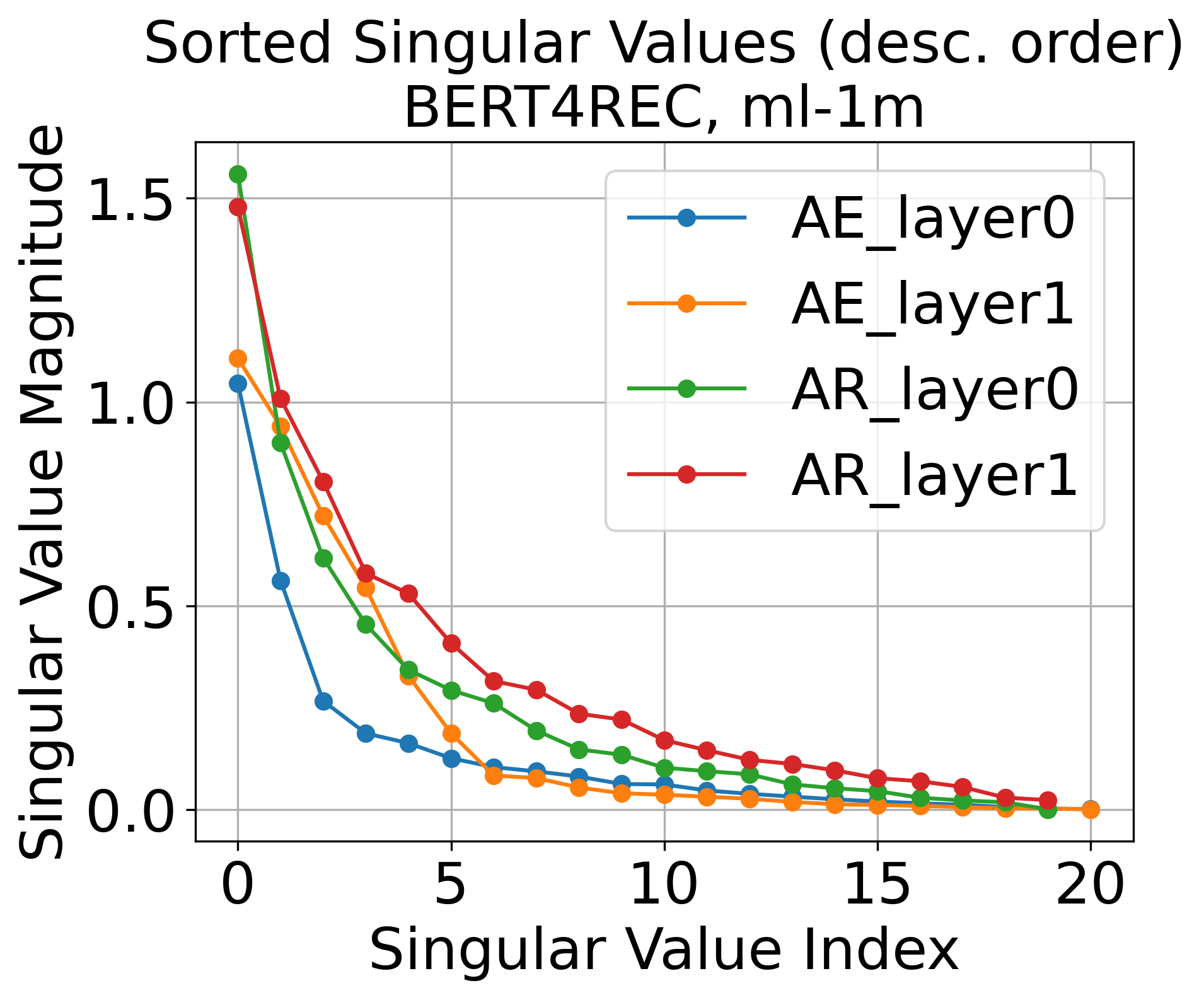}
        \caption{Vanilla attention, ML-1M}
        \label{fig:BERT4REC, ML-1M}
    \end{subfigure}
    \begin{subfigure}[]{0.49\linewidth}
        \centering
        \includegraphics[width=\textwidth]{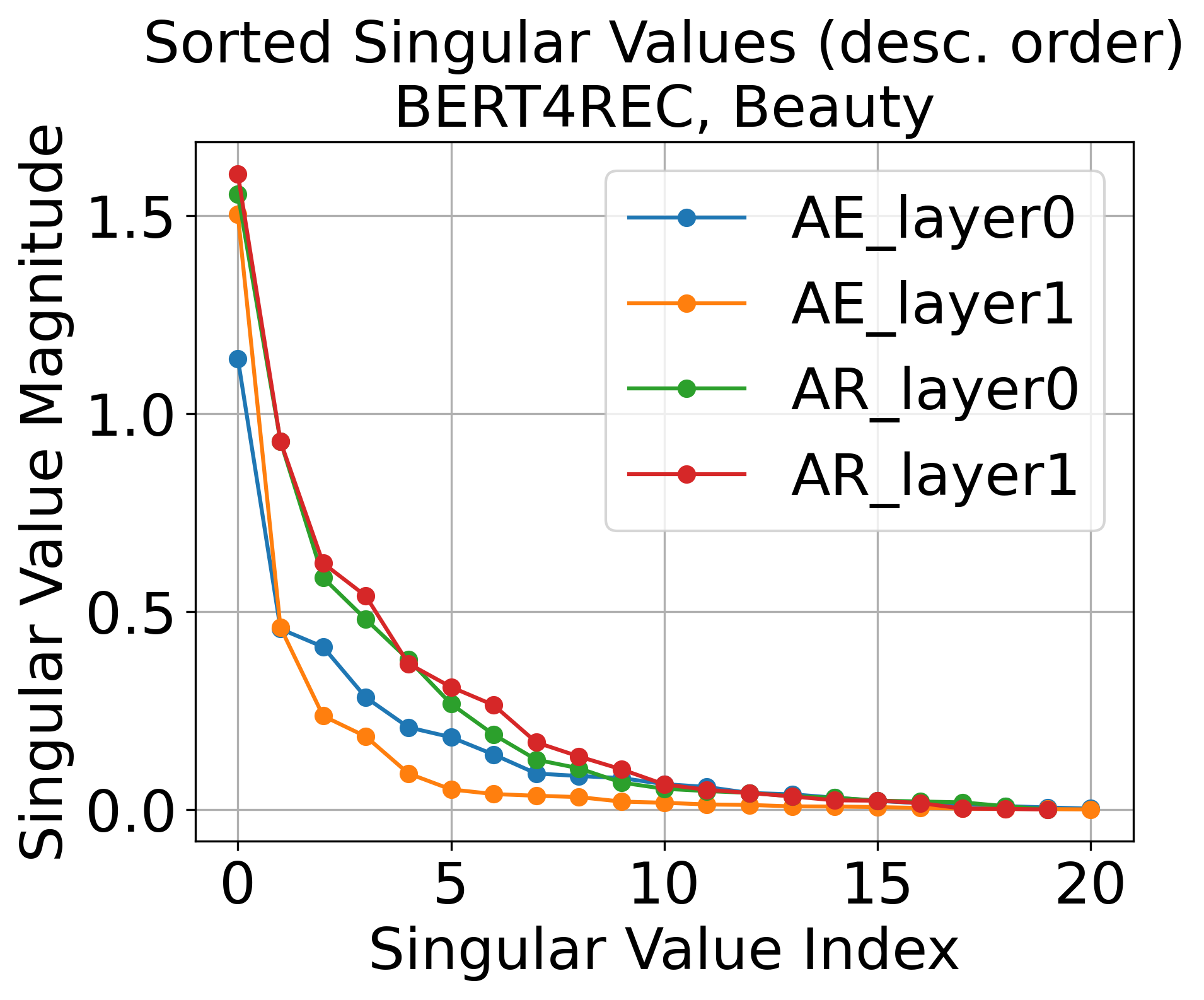}
        \caption{Vanilla attention, Beauty}
        \label{fig:BERT4REC, Beauty}
    \end{subfigure}
    \caption{Singular values distribution for random user X in descending order for Vanilla-\{AE, AR\} (BERT4Rec-like here), AE's singular values drops to near zero in a faster rate than AR, suggesting a lower rank approximation.}
    \label{fig:singular_value_show}
    \vspace{-10pt}
\end{figure}

%% file: 4_Large_scale_empirical_results.tex
\section{FINDING 3: Performance Report Supports AR's Theoretical Benefits}
\label{sec: emprical exp}
\input{figures/ModSAR.tex}
In this section, we focus on \textit{empirical performance} of AE and AR models to support our theoretical findings. On high level, we organize and execute a large-scale simulation of vanilla attention and their variants to reveal (1) an overall pattern of weather AE or AR attention performs better and (2) weather such performance superiority is robust across vanilla and variant attention modeling. Different than previous works \cite{petrov2022systematic} and \cite{klenitskiy2023turning}, we serve as the first work to run large scale and inclusive empirical experiments on comparing AE/AR models. In particular, we focus on three levels of AE/AR comparison:
\begin{enumerate}
    \item Vanilla AE/AR attention-base models comparison.
    \item Variant AE/AR attention-based models with locality, additional features and various loss functions.
    \item Variant AE/AR attention-based models of broader and more matured Huggingface community. 
\end{enumerate}

\subsection{Vanilla AE/AR Attention Comparison}
\label{sec: exp: vanilla attn}
\begin{table}[t]
\caption{Performance comparison of Vanilla-AE/AR self-attention. Best metrics are in bold. Vanilla-AR outperforms AE in 4/5 datasets by a large margin.}
\centering
\resizebox{\linewidth}{!}{%
\begin{tabular}{lccccccc}
\toprule
\textbf{Dataset}                 & \textbf{Attn.} & \textbf{R@5}    & \textbf{R@10}   & \textbf{R@20}   & \textbf{N@5}    & \textbf{N@10}   & \textbf{N@20}   \\ \midrule
\multirow{2}{*}{\textbf{Beauty}} & \textbf{AE}    & 0.0190          & 0.0302          & 0.0443          & 0.0122          & 0.0158          & 0.0193          \\
                                 & \textbf{AR}    & \textbf{0.0295} & \textbf{0.0391} & \textbf{0.0525} & \textbf{0.0220} & \textbf{0.0251} & \textbf{0.0284} \\ \midrule
\multirow{2}{*}{\textbf{Sports}} & \textbf{AE}    & 0.0102          & 0.0172          & 0.0279          & 0.0064          & 0.0087          & 0.0113          \\
                                 & \textbf{AR}    & \textbf{0.0175} & \textbf{0.0242} & \textbf{0.0334} & \textbf{0.0127} & \textbf{0.0148} & \textbf{0.0172} \\ \midrule
\multirow{2}{*}{\textbf{Video}}  & \textbf{AE}    & 0.0524          & 0.0837          & 0.1292          & 0.0337          & 0.0437          & 0.0552          \\
                                 & \textbf{AR}    & \textbf{0.0629} & \textbf{0.0941} & \textbf{0.1374} & \textbf{0.0422} & \textbf{0.0522} & \textbf{0.0631} \\ \midrule
\multirow{2}{*}{\textbf{Yelp}}   & \textbf{AE}    & \textbf{0.0397} & \textbf{0.0575} & \textbf{0.0861} & \textbf{0.0276} & \textbf{0.0333} & \textbf{0.0405} \\
                                 & \textbf{AR}    & 0.0254          & 0.0426          & 0.0694          & 0.0161          & 0.0216          & 0.0284          \\ \midrule
\multirow{2}{*}{\textbf{ML-1M}}  & \textbf{AE}    & 0.1490          & 0.2344          & 0.3425          & 0.0974          & 0.1248          & 0.1520          \\
                                 & \textbf{AR}    & \textbf{0.2086} & \textbf{0.3008} & \textbf{0.4116} & \textbf{0.1440} & \textbf{0.1738} & \textbf{0.2018} \\ \bottomrule
\end{tabular}%
}
\label{tab:prelim}
\end{table}
\subsubsection*{Findings} Empirically, Vanilla-AR outperforms Vanilla-AE attention overall on popular recommendation benchmarks.

We report the performance comparison of vanilla AE~/~AR attention in \Cref{tab:prelim}. Specific model definitions are introduced in \Cref{sec: prelim and overall setup}. \Cref{tab:prelim} shows that AR consistently outperforms AE in 4/5 datasets besides Yelp. The average performance gain of AR over AE is $25.35\%$ across all datasets and metrics, showing the benefit of AR models from its sparse local inductive bias and rich representation variance introduced in \Cref{tab: l1/l2 sparsity} and \Cref{sec: low rank approx}. The specific performance gains of Vanilla-AR over AE are 48.26\%, 58.79\%, 16.30\%, -31.33\%, 34.73\% for Beauty, Sports, Video, Yelp and ML-1M dataset; we also observe that the vanilla-AR exhibits superior ranking ability by showing a NDCG gain of 31.95\% than vanilla-AE. In contrast, the performance gain for Recall is lower as 18.75\%. We attribute the subpar performance of AR model on Yelp dataset to the underlying long-term visit effects. Based on these findings, we focus on comparing variant AE~/~AR models in the following sections.

\subsection{Variant AE/AR Attention Comparison under Various Design Choices}
\label{sec: exp common design space}
\input{tables/SAR}
\subsubsection*{Findings}
We show in the previous section that vanilla-AR outperforms AE overall. In this section, we are curious if this conclusion is robust in broader settings of more variant feature/model/loss designs and find: In the common design space for sequential recommendation, AR attention performs better than AE with variant attention types and loss types. Details are explained as follows.

\subsubsection{Experimental Setup}
\label{sec: common design choice exp set up}
In this section, we compare AE/AR with a focus on three aspects of attention variants, namely Feature, modeling and optimization choices:
\begin{enumerate}
    \item Feature: Incorporating user embeddings via SSE-PT.
    \item Modeling: Local attention (Local-attn).
    \item Optimization: Loss in \{Loss-CE, Loss-BPR\}.
\end{enumerate}
where \textit{SSE-PT} \cite{wu2020sse} is an effective way to combine user embeddings and item embeddings without causing overfitting; \textit{Local-attn} \cite{he2021locker} apply a window-based local attention mechanism in addition to vanilla self-attention. \textit{\{Loss-CE, Loss-BPR\}} serves as two mostly used loss functions for sequential recommendation; More details of comparison models are introduced in \Cref{sec: prelim and overall setup}. Due to computational resource limitations, we can not study every combination of designs but rather set one representative configuration for each aspect in \{Feature, Modeling, Optimization\}. For each of above three designs, we search the best hyperparameters for AE and AR respectively and compare their performance. To tune as many combinations of hyper-parameters with limited machines, we adopt Ray tune and Asynchronous Successive Halving (ASHA) heuristics for hyper-parameter tuning. ASHA dynamically compares each configuration and early-terminates bad-performing ones to save resources. The entire experimental pipeline is named \underline{\textbf{Mo}}dularized \underline{\textbf{D}}esign Space for \underline{\textbf{S}}elf-\underline{\textbf{A}}ttentive \underline{\textbf{R}}ecommender (\textbf{ModSAR}) shown in \Cref{fig:ModSAR}. Evaluation results for different design options are discussed in section \ref{sar results_}. AE~/~AR models of each design choice are derived by adding/removing the causal mask in the attention module.

\subsubsection{Experimental Results}
\label{sar results_}
Table \ref{tab:SAR} shows AE/AR comparison based on our common design space with adaptive hyperparameter search. We report the evaluation metrics and relative improvements as percentages of the best AR models over AE ones in bold. The better model in \{AE, AR\} of each configuration in \{Local-attn, SSE-PT, Loss-CE, Loss-BPR\} is underlined; AR on average outperforms AE by a large margin of 19.64\%, 24.01\%, 28.04\%, 35.77\% for Local-attn, SSE-PE, Loss-CE, and Loss-BPR. The best AR model of all configurations on average leads the best AE model by 21.60\%. This shows that (1) the concept of auto-regression (AR) continues to perform better than AE under extensive experimentation of common design space with well-searched hyperparameters. (2) AR's performance gain also varies on different data distributions. AR leads AE by 39.73\%, 26.01\%, 32.46\%, 8.28\%, 27.84\% for Beauty, Sports, Video, Yelp and ML-1M datasets, showing that AR's robustness for both sparse and dense datasets besides Yelp due to its long-term visiting effects for local business. (3) AR models continue to show better ranking performance than AE, as AR's average NDCG gain over AE is 29.69\% across all datasets, models and metrics and the gain for Recall is around 5\% lower as 24.04\%. (5) Compared with results from section \ref{sec: exp: vanilla attn}, we found that AR's superior performance is retained here when expanding the model variants in common design space (performance gain of AR 25.35\% in \Cref{sec: exp: vanilla attn}, and 21.60\% here).

In summary, AR continues to outperform AE across various design choices; AR outperforms AE on both sparse and dense datasets in our common design space; AR continues to show better ranking ability than AE as we showed in point (3).

\subsection{Variant AE/AR Attention Comparison with Advanced Models from Huggingface}
\label{sec: exp huggingface}
\subsubsection*{Findings} In this section, we are curious whether AE or AR performs better when scaling our research objective to a broader group of advanced self-attention variants in Huggingface. Thus, we design a HuggingFace plug-in that directly applies AE~/~AR-based language models to recommendation datasets (An item ID sequence as a "sentence"). We find that AR attention continues to outperform AE when directly applying and tuning advanced Huggingface models on recommendation datasets. Details are explained as follows.
\label{sec:HF}
\input{tables/HF}
\subsubsection{Experimental Setup}
Bridging the gap between the Sequential Recommenders and fast-developing NLP models has been long talked about \cite{de2021transformers4rec}. Here, we focus on further extensive study of AE and AR comparison on the latest NLP models from HuggingFace. Specifically, we select three popular models that are originally uni-directional (Llama, Transformer-XL) or bi-directional (ALBERT). Similar to the approach discussed in \Cref{sec: exp common design space}, we change the original models to their AE/AR counterparts by (1) adding/removing the causal mask and (2) changing their optimization task to the next item prediction/Mask Language Model (MLM). We keep here the adaptive hyperparameter search introduced in \Cref{sec: common design choice exp set up}. Evaluation results for HuggingFace models are discussed in \Cref{results hf}.

\subsubsection{Experimental Results}
\label{results hf}
Table \ref{tab:HF} shows the evaluation results of the above models and the relative improvement of the best AR models over AE models in bold. The better of \{AE, AR\} for each model choice is underlined. According to table \ref{tab:HF}, (1) AR continues to outperform AE on average by 16.21\%, 14.37\%, 15.30\% for ALBERT Llama and Transformer-XL; The best AR model outperforms the best AE model by 16.00\%. (2) AR performs extra well on sparse datasets. For instance, AR on average leads AE by 35.80\%, 21.63\% for Beauty and Sports which are top-2 sparsest datasets, and by 14.42\%, 17.67\% for Video and ML-1M which are bottom-2 sparsest. AR only loses to AE on Yelp due to its intrinsic long-term visiting effects of local businesses. (3) AR continues to show superior ranking ability than AE. This is shown by the fact that AR outperforms AE by 12.46\% for Recall and by a higher value of 18.12\% for NDCG.

In summary, AR continues to show superior performance than AE on advanced HuggingFace NLP models, this is consistent with our conclusions from \Cref{sec: exp: vanilla attn} and \Cref{sec: exp common design space}. AR's performance gain is more significant on sparse datasets. AR shows better ranking ability than AE by having a more improved NDCG than Recall.

\section{Connecting Empirical Results with Theoretical Analysis}
\label{sec: connect empirical and theory}
\begin{figure}[t]
    \centering
    \begin{subfigure}[]{0.49\linewidth}
        \centering
        \includegraphics[width=\textwidth]{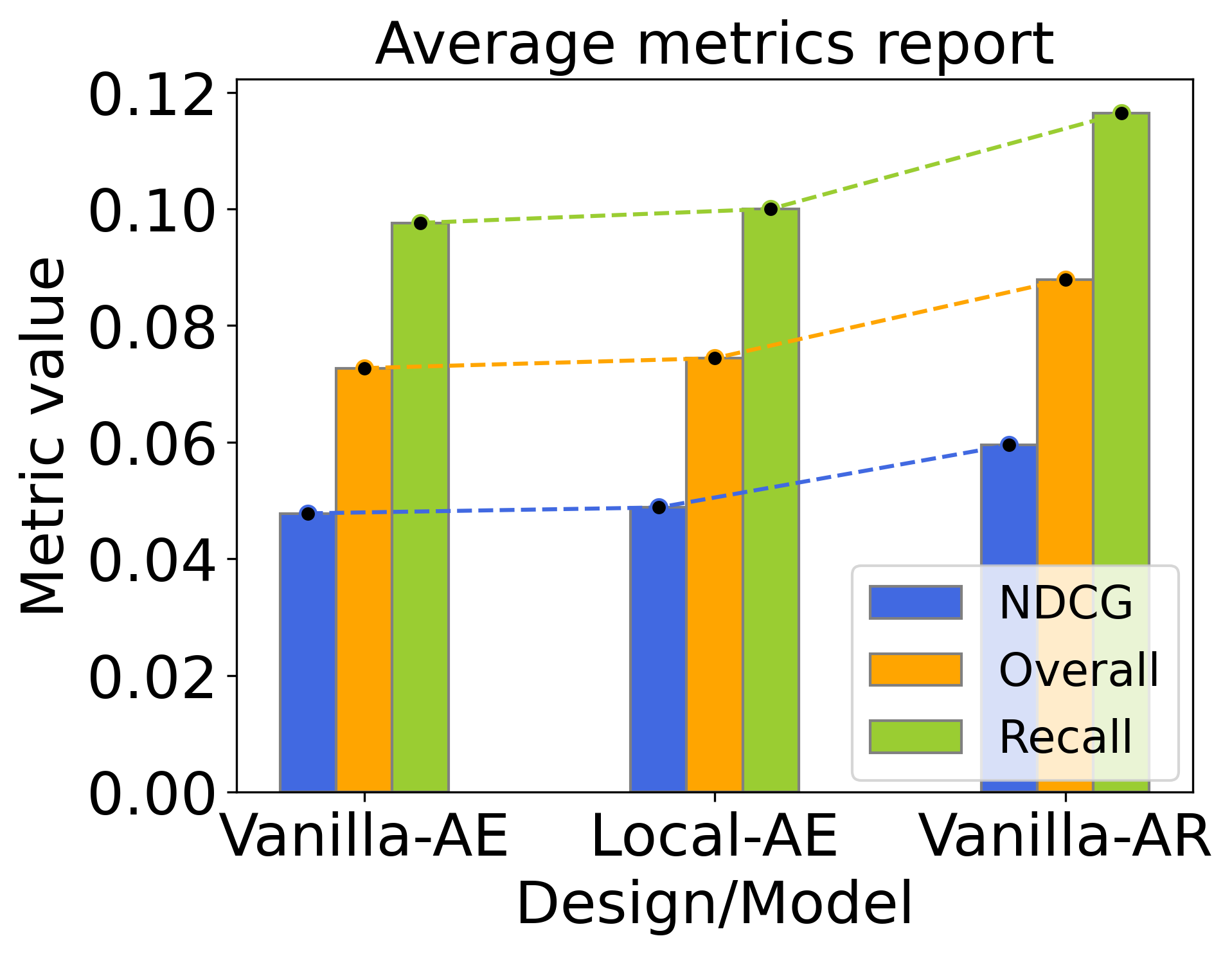}
        \caption{Vanilla and Local-AE models.}
        \label{fig:bar_locality}
    \end{subfigure}
    \begin{subfigure}[]{0.47\linewidth}
        \centering
        \includegraphics[width=\textwidth]{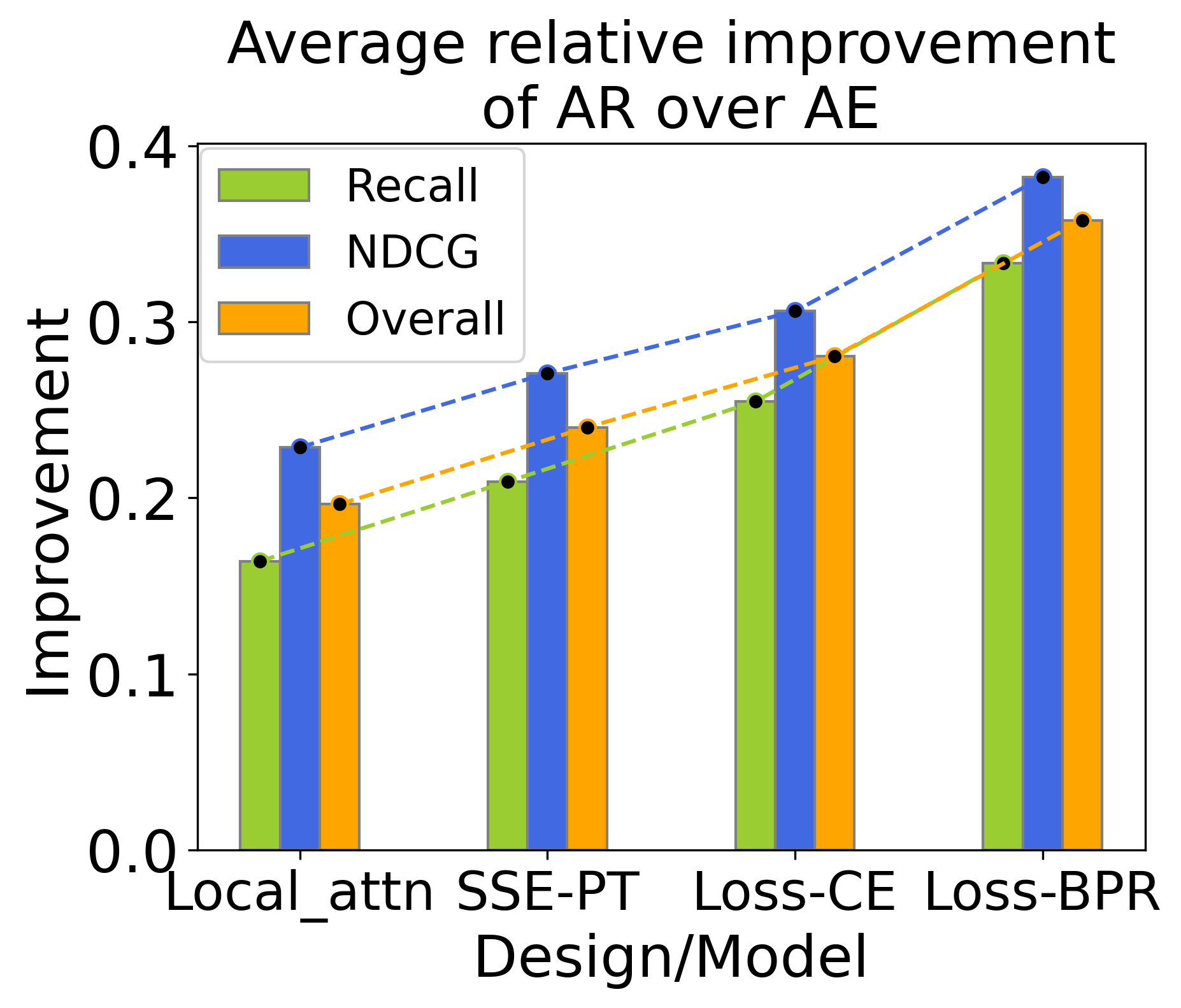}
        \caption{Common design space.}
        \label{fig:bar_SAR}
    \end{subfigure}
    \caption{\Cref{fig:bar_locality} shows the average performance increases with growing local inductive bias consistent with our theoretical analysis. Local-attn in \Cref{fig:bar_SAR} shows lowest improvements due to already injected inductive bias in Local-attn-AE.}
    \label{fig: performance visual}
    \vspace{-10pt}
\end{figure}

As proposed in \Cref{sec: sparsity measure}, stronger sparse local inductive bias and neighborhood effects facilitate model performance. Thus, we visualize in figure \Cref{fig:bar_locality} the average performance of Vanilla-AR, Local-AE and Vanilla-AR as the neighborhood effect increases from left to right. \Cref{fig:bar_locality} shows that the average performance indeed increases as locality grows stronger. This justifies our theoretical analysis emphasizing the benefits of sparse local inductive bias. We also visualize in \Cref{fig:bar_SAR} the average relative performance improvement of AR over AE for models in our common design choice. \Cref{fig:bar_SAR} shows that Local-attn has the lowest AR improvement over AE. We attribute this phenomenon to the already existing local inductive bias in the Local-attn-AE model where attention scores are constrained to a local window (also see the middle sub-figure in \Cref{fig:attn_visual_1row}). Overall, our empirical results show alignment with theoretical analysis highlighting locality/neighborhood advantages of AR in \Cref{sec: sparsity measure} and \Cref{sec: low rank approx}.

\section{Conclusion}
\label{sec: conclusion}
In previous sections, we have conducted a theoretical analysis of AE/AR attention from two perspectives, namely (1) sparse local inductive bias and (2) low-rank approximation. We highlight the sparse neighborhood effects for the AR attention as the performance contributor shown in \Cref{fig:attn_visual}. This is further justified by our empirical results that performance increases with stronger locality/neighborhood effects shown in \Cref{fig: performance visual}. Additionally, we find that the AR attention stores more variant data dynamics as it requires a higher-rank SVD approximation determined by the noise-based hard-thresholding method introduced in \Cref{sec: low rank approx}; further extensive experiments have also demonstrated AR's robust performance superiority over AE for both vanilla and variant self-attentive models. We conclude here that AR models serve as a more reasonable starting point when developing future sequential recommenders given their desired mathematical properties and superior performance. More efforts should be exerted on behavior analysis of locality-based attention and related variants to unlock the potentials of such model designs.

%% file: figures/ModSAR.tex
\begin{figure}[t]
  \centering
  \includegraphics[width=1.0\linewidth]{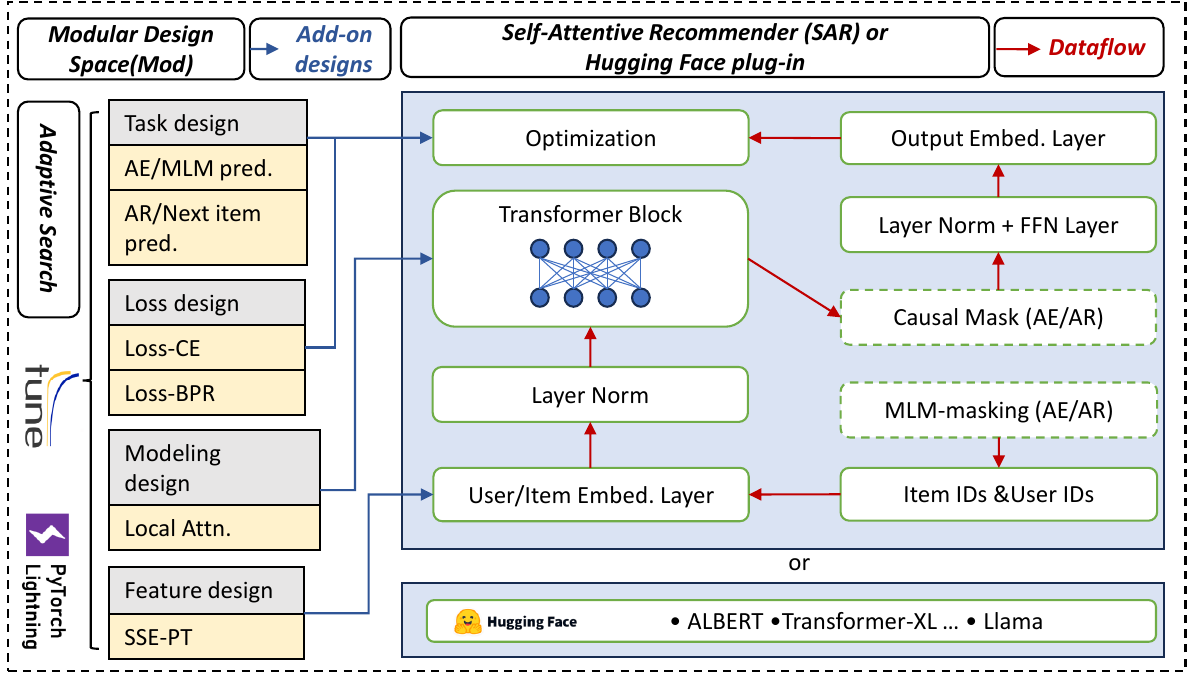}
  \caption{The overall architecture for our ModSAR. The right part introduces the self-attentive backbone controlled by the left part of our modularized design space of \{Feature, Modeling, Loss and Task\}. User can also choose between \{self-attention + design space, Huggingface models\}. Ray tune and ASHA-adaptive search are utilized to manage experiments.}
  \label{fig:ModSAR}
  \vspace{-10pt}
\end{figure}

%% file: tables/SAR.tex
\begin{table}[t]
\small
\caption{Performance comparison of variant AE~/~AR model under common design space. The better model in \{AE, AR\} are underlined for each design choice. Improvements are based on the best AE/AR model in bold. AR outperforms AE at least on 4/5 datasets for all design choices. The best AR model sometimes significantly outperforms the best AE one (often over 20\%).}
\resizebox{\linewidth}{!}{%
\begin{tabular}{@{}llccccccccc@{}}
\toprule
\multicolumn{2}{c}{\textbf{Design}}                      & \textbf{Local} & \textbf{Local}   & \textbf{SSE} & \textbf{SSE} & \textbf{Loss}      & \textbf{Loss}      & \textbf{Loss} & \textbf{Loss} & \textbf{Best}   \\
\multicolumn{2}{c}{\textbf{Choice}}                      & \textbf{attn} & \textbf{attn}   & \textbf{PT} & \textbf{PT} & \textbf{CE}      & \textbf{CE}      & \textbf{BPR} & \textbf{BPR} & \textbf{Improv.}    \\ \midrule
\textbf{Data}                 & \textbf{-} & \textbf{AE}         & \textbf{AR}           & \textbf{AE}     & \textbf{AR}     & \textbf{AE}           & \textbf{AR}           & \textbf{AE}       & \textbf{AR}       & \textbf{$\frac{AR-AE}{AE}$} \\ \midrule
\multirow{6}{*}{\textbf{Bea.}} & R@5             & \textbf{.0234}     & {\ul \textbf{.0324}} & .0172          & {\ul .0239}    & .0203                & {\ul .0313}          & .0095            & {\ul .0133}      & 38.46\%             \\
                                 & R@10            & \textbf{.0371}     & {\ul \textbf{.0477}} & .0280          & {\ul .0358}    & .0330                & {\ul .0468}          & .0168            & {\ul .0237}      & 28.57\%             \\
                                 & R@20            & \textbf{.0544}     & {\ul .0662}          & .0430          & {\ul .0512}    & .0496                & {\ul \textbf{.0673}} & .0279            & {\ul .0384}      & 23.71\%             \\
                                 & N@5               & \textbf{.0136}     & {\ul \textbf{.0206}} & .0105          & {\ul .0152}    & .0116                & {\ul .0192}          & .0054            & {\ul .0076}      & 51.47\%             \\
                                 & N@10              & \textbf{.0180}     & {\ul \textbf{.0255}} & .0140          & {\ul .0190}    & .0157                & {\ul .0242}          & .0077            & {\ul .0110}      & 41.67\%             \\
                                 & N@20              & \textbf{.0223}     & {\ul \textbf{.0302}} & .0177          & {\ul .0229}    & .0199                & {\ul .0294}          & .0105            & {\ul .0146}      & 35.43\%             \\ \midrule
\multirow{6}{*}{\textbf{Spo.}} & R@5             & \textbf{.0141}     & {\ul .0184}          & .0124          & {\ul .0151}    & .0132                & {\ul \textbf{.0188}} & .0071            & {\ul .0088}      & 33.33\%             \\
                                 & R@10            & \textbf{.0238}     & {\ul .0286}          & .0192          & {\ul .0228}    & .0216                & {\ul \textbf{.0288}} & .0122            & {\ul .0150}      & 21.01\%             \\
                                 & R@20            & \textbf{.0374}     & {\ul .0414}          & .0288          & {\ul .0335}    & .0349                & {\ul \textbf{.0429}} & .0199            & {\ul .0243}      & 14.71\%             \\
                                 & N@5               & \textbf{.0082}     & {\ul .0111}          & .0075          & {\ul .0096}    & .0078                & {\ul \textbf{.0114}} & .0042            & {\ul .0051}      & 39.02\%             \\
                                 & N@10              & \textbf{.0113}     & {\ul .0144}          & .0097          & {\ul .0121}    & .0105                & {\ul \textbf{.0146}} & .0058            & {\ul .0071}      & 29.20\%             \\
                                 & N@20              & \textbf{.0147}     & {\ul .0176}          & .0121          & {\ul .0148}    & .0138                & {\ul \textbf{.0181}} & .0078            & {\ul .0094}      & 23.13\%             \\ \midrule
\multirow{6}{*}{\textbf{Vid.}}  & R@5             & \textbf{.0534}     & {\ul .0681}          & .0444          & {\ul .0551}    & .0500                & {\ul \textbf{.0692}} & .0293            & {\ul .0470}      & 29.59\%             \\
                                 & R@10            & \textbf{.0872}     & {\ul .1076}          & .0744          & {\ul .0890}    & .0844                & {\ul \textbf{.1125}} & .0548            & {\ul .0777}      & 29.01\%             \\
                                 & R@20            & \textbf{.1350}     & {\ul .1571}          & .1157          & {\ul .1355}    & .1304                & {\ul \textbf{.1664}} & .0892            & {\ul .1235}      & 23.26\%             \\
                                 & N@5               & \textbf{.0321}     & {\ul .0413}          & .0266          & {\ul .0332}    & .0296                & {\ul \textbf{.0421}} & .0173            & {\ul .0280}      & 31.15\%             \\
                                 & N@10              & \textbf{.0430}     & {\ul .0540}          & .0362          & {\ul .0441}    & .0406                & {\ul \textbf{.0560}} & .0255            & {\ul .0379}      & 30.23\%             \\
                                 & N@20              & \textbf{.0550}     & {\ul .0665}          & .0465          & {\ul .0559}    & .0522                & {\ul \textbf{.0696}} & .0341            & {\ul .0494}      & 26.55\%             \\ \midrule
\multirow{6}{*}{\textbf{Yel.}}   & R@5             & {\ul .0422}        & .0386                & .0298          & {\ul .0337}    & {\ul \textbf{.0504}} & \textbf{.0477}       & .0244            & {\ul .0324}      & -5.36\%             \\
                                 & R@10            & {\ul .0649}        & .0592                & .0490          & {\ul .0527}    & {\ul \textbf{.0728}} & \textbf{.0724}       & .0423            & {\ul .0524}      & -0.55\%             \\
                                 & R@20            & {\ul .0992}        & .0922                & .0762          & {\ul .0807}    & {\ul \textbf{.1057}} & \textbf{.1057}       & .0695            & {\ul .0825}      & 0.00\%              \\
                                 & N@5               & {\ul .0272}        & .0263                & .0190          & {\ul .0220}    & {\ul \textbf{.0353}} & \textbf{.0323}       & .0140            & {\ul .0210}      & -8.50\%             \\
                                 & N@10              & {\ul .0345}        & .0329                & .0252          & {\ul .0281}    & {\ul \textbf{.0424}} & \textbf{.0402}       & .0197            & {\ul .0274}      & -5.19\%             \\
                                 & N@20              & {\ul .0431}        & .0412                & .0321          & {\ul .0351}    & {\ul \textbf{.0507}} & \textbf{.0486}       & .0266            & {\ul .0350}      & -4.14\%             \\ \midrule
\multirow{6}{*}{\textbf{ML-.}}  & R@5             & \textbf{.1715}     & {\ul \textbf{.2161}} & .1232          & {\ul .1745}    & .1641                & {\ul .2096}          & .1116            & {\ul .1591}      & 26.01\%             \\
                                 & R@10            & \textbf{.2666}     & {\ul \textbf{.3118}} & .2040          & {\ul .2523}    & .2606                & {\ul .3089}          & .1868            & {\ul .2449}      & 16.95\%             \\
                                 & R@20            & \textbf{.3892}     & {\ul \textbf{.4290}} & .2942          & {\ul .3487}    & .3724                & {\ul .4174}          & .2960            & {\ul .3629}      & 10.23\%             \\
                                 & N@5               & \textbf{.1057}     & {\ul \textbf{.1362}} & .0749          & {\ul .1127}    & .0991                & {\ul .1316}          & .0671            & {\ul .0966}      & 28.86\%             \\
                                 & N@10              & \textbf{.1363}     & {\ul \textbf{.1671}} & .1009          & {\ul .1377}    & .1301                & {\ul .1638}          & .0912            & {\ul .1243}      & 22.60\%             \\
                                 & N@20              & \textbf{.1671}     & {\ul \textbf{.1966}} & .1237          & {\ul .1619}    & .1582                & {\ul .1912}          & .1186            & {\ul .1540}      & 17.65\%             \\ \bottomrule
\end{tabular}%
}
\label{tab:SAR}
\vspace{-15pt}
\end{table}

%% file: tables/HF.tex
\begin{table}[t]
\footnotesize
\caption{Performance comparison of AE~/~AR with advanced models in broader Huggingface AI community. The better model in \{AE, AR\} are underlined. Improvements are based on the best AE~/~AR model in bold. AR outperforms AE on at least 4/5 datasets. The best AR model sometimes significantly outperforms the best AE model (often over 20\% for \textit{Beauty} and \textit{Sports}, over around 10\% for \textit{Video} and \textit{ML-1M}).}
\resizebox{\linewidth}{!}{%
\begin{tabular}{llccccccc}
\toprule
\multicolumn{2}{c}{\textbf{HF}}                      & \textbf{AL-} & \textbf{AL-} & \textbf{LLa-}        & \textbf{LLa-}        & \textbf{Trm.-}  & \textbf{Trm.-}  & \textbf{Best}    \\
\multicolumn{2}{c}{\textbf{Model}}                      & \textbf{BERT} & \textbf{BERT} & \textbf{MA}        & \textbf{MA}        & \textbf{XL}  & \textbf{XL}  & \textbf{Improv.}    \\ \midrule
\textbf{Data}                 & \textbf{-} & \textbf{AE}     & \textbf{AR}     & \textbf{AE}           & \textbf{AR}           & \textbf{AE}           & \textbf{AR}           & \textbf{$\frac{AR-AE}{AE}$} \\ \midrule
\multirow{6}{*}{\textbf{Bea.}} & R@5             & .0189          & {\ul .0284}    & .0192                & {\ul .0277}          & \textbf{.0247}       & {\ul \textbf{.0343}} & 38.87\%             \\
                                 & R@10            & .0312          & {\ul .0386}    & .0319                & {\ul .0394}          & \textbf{.0389}       & {\ul \textbf{.0485}} & 24.68\%             \\
                                 & R@20            & .0481          & {\ul .0527}    & .0470                & {\ul .0540}          & \textbf{.0553}       & {\ul \textbf{.0666}} & 20.43\%             \\
                                 & N@5               & .0113          & {\ul .0192}    & .0118                & {\ul .0180}          & \textbf{.0150}       & {\ul \textbf{.0226}} & 50.67\%             \\
                                 & N@10              & .0152          & {\ul .0224}    & .0159                & {\ul .0218}          & \textbf{.0195}       & {\ul \textbf{.0272}} & 39.49\%             \\
                                 & N@20              & .0195          & {\ul .0260}    & .0197                & {\ul .0255}          & \textbf{.0236}       & {\ul \textbf{.0317}} & 34.32\%             \\ \midrule
\multirow{6}{*}{\textbf{Spo.}} & R@5             & .0135          & {\ul .0160}    & .0124                & {\ul .0154}          & \textbf{.0146}       & {\ul \textbf{.0187}} & 28.08\%             \\
                                 & R@10            & .0216          & {\ul .0238}    & .0199                & {\ul .0233}          & \textbf{.0230}       & {\ul \textbf{.0294}} & 27.83\%             \\
                                 & R@20            & .0347          & {\ul .0351}    & .0309                & {\ul .0338}          & \textbf{.0359}       & {\ul \textbf{.0432}} & 20.33\%             \\
                                 & N@5               & .0079          & {\ul .0101}    & .0073                & {\ul .0098}          & \textbf{.0088}       & {\ul \textbf{.0119}} & 35.23\%             \\
                                 & N@10              & .0106          & {\ul .0125}    & .0098                & {\ul .0124}          & \textbf{.0115}       & {\ul \textbf{.0153}} & 33.04\%             \\
                                 & N@20              & .0139          & {\ul .0154}    & .0125                & {\ul .0150}          & \textbf{.0148}       & {\ul \textbf{.0188}} & 27.03\%             \\ \midrule
\multirow{6}{*}{\textbf{Vid.}}  & R@5             & .0471          & {\ul .0625}    & .0556                & {\ul .0625}          & \textbf{.0644}       & {\ul \textbf{.0709}} & 10.09\%             \\
                                 & R@10            & .0798          & {\ul .0955}    & .0914                & {\ul .0980}          & \textbf{.1032}       & {\ul \textbf{.1106}} & 7.17\%              \\
                                 & R@20            & .1238          & {\ul .1420}    & .1407                & {\ul .1436}          & \textbf{.1543}       & {\ul \textbf{.1620}} & 4.99\%              \\
                                 & N@5               & .0284          & {\ul .0388}    & .0339                & {\ul .0387}          & \textbf{.0391}       & {\ul \textbf{.0441}} & 12.79\%             \\
                                 & N@10              & .0389          & {\ul .0494}    & .0454                & {\ul .0501}          & \textbf{.0516}       & {\ul \textbf{.0569}} & 10.27\%             \\
                                 & N@20              & .0500          & {\ul .0611}    & .0577                & {\ul .0616}          & \textbf{.0645}       & {\ul \textbf{.0699}} & 8.37\%              \\ \midrule
\multirow{6}{*}{\textbf{Yel.}}   & R@5             & {\ul .0329}    & .0219          & {\ul .0364}          & \textbf{.0347}       & {\ul \textbf{.0377}} & .0343                & -7.96\%             \\
                                 & R@10            & {\ul .0514}    & .0392          & {\ul .0557}          & .0529                & \textbf{.0563}       & {\ul \textbf{.0567}} & 0.71\%              \\
                                 & R@20            & {\ul .0798}    & .0632          & {\ul \textbf{.0865}} & .0791                & .0845                & {\ul \textbf{.0903}} & 4.39\%              \\
                                 & N@5               & {\ul .0217}    & .0128          & .0237                & {\ul \textbf{.0246}} & {\ul \textbf{.0259}} & .0210                & -5.02\%             \\
                                 & N@10              & {\ul .0276}    & .0184          & .0299                & {\ul \textbf{.0305}} & {\ul \textbf{.0319}} & .0282                & -4.39\%             \\
                                 & N@20              & {\ul .0348}    & .0244          & {\ul .0377}          & \textbf{.0370}       & {\ul \textbf{.0390}} & .0365                & -5.13\%             \\ \midrule
\multirow{6}{*}{\textbf{ML-.}}  & R@5             & .1508          & {\ul .2108}    & .1733                & {\ul .2013}          & \textbf{.1988}       & {\ul \textbf{.2238}} & 12.58\%             \\
                                 & R@10            & .2371          & {\ul .3018}    & .2714                & {\ul .2894}          & \textbf{.2935}       & {\ul \textbf{.3219}} & 9.68\%              \\
                                 & R@20            & .3475          & {\ul .4081}    & .3894                & {\ul .4017}          & \textbf{.4000}       & {\ul \textbf{.4311}} & 7.77\%              \\
                                 & N@5               & .0908          & {\ul .1322}    & .1106                & {\ul .1282}          & \textbf{.1250}       & {\ul \textbf{.1412}} & 12.96\%             \\
                                 & N@10              & .1185          & {\ul .1615}    & .1422                & {\ul .1565}          & \textbf{.1556}       & {\ul \textbf{.1728}} & 11.05\%             \\
                                 & N@20              & .1464          & {\ul .1883}    & .1722                & {\ul .1849}          & \textbf{.1827}       & {\ul \textbf{.2003}} & 9.63\%              \\ \bottomrule
\end{tabular}%
}
\label{tab:HF}
\vspace{-10pt}
\end{table}

%% file: main.bbl

\begin{thebibliography}{40}


\ifx \showCODEN    \undefined \def \showCODEN     #1{\unskip}     \fi
\ifx \showDOI      \undefined \def \showDOI       #1{#1}\fi
\ifx \showISBNx    \undefined \def \showISBNx     #1{\unskip}     \fi
\ifx \showISBNxiii \undefined \def \showISBNxiii  #1{\unskip}     \fi
\ifx \showISSN     \undefined \def \showISSN      #1{\unskip}     \fi
\ifx \showLCCN     \undefined \def \showLCCN      #1{\unskip}     \fi
\ifx \shownote     \undefined \def \shownote      #1{#1}          \fi
\ifx \showarticletitle \undefined \def \showarticletitle #1{#1}   \fi
\ifx \showURL      \undefined \def \showURL       {\relax}        \fi
\providecommand\bibfield[2]{#2}
\providecommand\bibinfo[2]{#2}
\providecommand\natexlab[1]{#1}
\providecommand\showeprint[2][]{arXiv:#2}

\bibitem[Brown et~al\mbox{.}(2020)]%
        {brown2020language}
\bibfield{author}{\bibinfo{person}{Tom Brown}, \bibinfo{person}{Benjamin Mann}, \bibinfo{person}{Nick Ryder}, \bibinfo{person}{Melanie Subbiah}, \bibinfo{person}{Jared~D Kaplan}, \bibinfo{person}{Prafulla Dhariwal}, \bibinfo{person}{Arvind Neelakantan}, \bibinfo{person}{Pranav Shyam}, \bibinfo{person}{Girish Sastry}, \bibinfo{person}{Amanda Askell}, {et~al\mbox{.}}} \bibinfo{year}{2020}\natexlab{}.
\newblock \showarticletitle{Language models are few-shot learners}.
\newblock \bibinfo{journal}{\emph{Advances in neural information processing systems}}  \bibinfo{volume}{33} (\bibinfo{year}{2020}), \bibinfo{pages}{1877--1901}.
\newblock


\bibitem[Cai et~al\mbox{.}(2010)]%
        {cai2010singular}
\bibfield{author}{\bibinfo{person}{Jian-Feng Cai}, \bibinfo{person}{Emmanuel~J Cand{\`e}s}, {and} \bibinfo{person}{Zuowei Shen}.} \bibinfo{year}{2010}\natexlab{}.
\newblock \showarticletitle{A singular value thresholding algorithm for matrix completion}.
\newblock \bibinfo{journal}{\emph{SIAM Journal on optimization}} \bibinfo{volume}{20}, \bibinfo{number}{4} (\bibinfo{year}{2010}), \bibinfo{pages}{1956--1982}.
\newblock


\bibitem[Candes and Recht(2012)]%
        {candes2012exact}
\bibfield{author}{\bibinfo{person}{Emmanuel Candes} {and} \bibinfo{person}{Benjamin Recht}.} \bibinfo{year}{2012}\natexlab{}.
\newblock \showarticletitle{Exact matrix completion via convex optimization}.
\newblock \bibinfo{journal}{\emph{Commun. ACM}} \bibinfo{volume}{55}, \bibinfo{number}{6} (\bibinfo{year}{2012}), \bibinfo{pages}{111--119}.
\newblock


\bibitem[Cho et~al\mbox{.}(2020)]%
        {cho2020meantime}
\bibfield{author}{\bibinfo{person}{Sung~Min Cho}, \bibinfo{person}{Eunhyeok Park}, {and} \bibinfo{person}{Sungjoo Yoo}.} \bibinfo{year}{2020}\natexlab{}.
\newblock \showarticletitle{MEANTIME: Mixture of attention mechanisms with multi-temporal embeddings for sequential recommendation}. In \bibinfo{booktitle}{\emph{Proceedings of the 14th ACM Conference on recommender systems}}. \bibinfo{pages}{515--520}.
\newblock


\bibitem[de~Souza Pereira~Moreira et~al\mbox{.}(2021)]%
        {de2021transformers4rec}
\bibfield{author}{\bibinfo{person}{Gabriel de Souza Pereira~Moreira}, \bibinfo{person}{Sara Rabhi}, \bibinfo{person}{Jeong~Min Lee}, \bibinfo{person}{Ronay Ak}, {and} \bibinfo{person}{Even Oldridge}.} \bibinfo{year}{2021}\natexlab{}.
\newblock \showarticletitle{Transformers4rec: Bridging the gap between nlp and sequential/session-based recommendation}. In \bibinfo{booktitle}{\emph{Proceedings of the 15th ACM Conference on Recommender Systems}}. \bibinfo{pages}{143--153}.
\newblock


\bibitem[Devlin et~al\mbox{.}(2018)]%
        {devlin2018bert}
\bibfield{author}{\bibinfo{person}{Jacob Devlin}, \bibinfo{person}{Ming-Wei Chang}, \bibinfo{person}{Kenton Lee}, {and} \bibinfo{person}{Kristina Toutanova}.} \bibinfo{year}{2018}\natexlab{}.
\newblock \showarticletitle{Bert: Pre-training of deep bidirectional transformers for language understanding}.
\newblock \bibinfo{journal}{\emph{arXiv preprint arXiv:1810.04805}} (\bibinfo{year}{2018}).
\newblock


\bibitem[Fan et~al\mbox{.}(2021)]%
        {fan2021lighter}
\bibfield{author}{\bibinfo{person}{Xinyan Fan}, \bibinfo{person}{Zheng Liu}, \bibinfo{person}{Jianxun Lian}, \bibinfo{person}{Wayne~Xin Zhao}, \bibinfo{person}{Xing Xie}, {and} \bibinfo{person}{Ji-Rong Wen}.} \bibinfo{year}{2021}\natexlab{}.
\newblock \showarticletitle{Lighter and better: low-rank decomposed self-attention networks for next-item recommendation}. In \bibinfo{booktitle}{\emph{Proceedings of the 44th international ACM SIGIR conference on research and development in information retrieval}}. \bibinfo{pages}{1733--1737}.
\newblock


\bibitem[Gao et~al\mbox{.}(2018)]%
        {gao2018low}
\bibfield{author}{\bibinfo{person}{Pengzhi Gao}, \bibinfo{person}{Ren Wang}, \bibinfo{person}{Meng Wang}, {and} \bibinfo{person}{Joe~H Chow}.} \bibinfo{year}{2018}\natexlab{}.
\newblock \showarticletitle{Low-rank matrix recovery from noisy, quantized, and erroneous measurements}.
\newblock \bibinfo{journal}{\emph{IEEE Transactions on Signal Processing}} \bibinfo{volume}{66}, \bibinfo{number}{11} (\bibinfo{year}{2018}), \bibinfo{pages}{2918--2932}.
\newblock


\bibitem[He et~al\mbox{.}(2017)]%
        {he2017neural}
\bibfield{author}{\bibinfo{person}{Xiangnan He}, \bibinfo{person}{Lizi Liao}, \bibinfo{person}{Hanwang Zhang}, \bibinfo{person}{Liqiang Nie}, \bibinfo{person}{Xia Hu}, {and} \bibinfo{person}{Tat-Seng Chua}.} \bibinfo{year}{2017}\natexlab{}.
\newblock \showarticletitle{Neural collaborative filtering}. In \bibinfo{booktitle}{\emph{Proceedings of the 26th international conference on world wide web}}. \bibinfo{pages}{173--182}.
\newblock


\bibitem[He et~al\mbox{.}(2021)]%
        {he2021locker}
\bibfield{author}{\bibinfo{person}{Zhankui He}, \bibinfo{person}{Handong Zhao}, \bibinfo{person}{Zhe Lin}, \bibinfo{person}{Zhaowen Wang}, \bibinfo{person}{Ajinkya Kale}, {and} \bibinfo{person}{Julian McAuley}.} \bibinfo{year}{2021}\natexlab{}.
\newblock \showarticletitle{Locker: Locally constrained self-attentive sequential recommendation}. In \bibinfo{booktitle}{\emph{Proceedings of the 30th ACM International Conference on Information \& Knowledge Management}}. \bibinfo{pages}{3088--3092}.
\newblock


\bibitem[Hurley and Rickard(2009)]%
        {hurley2009comparing}
\bibfield{author}{\bibinfo{person}{Niall Hurley} {and} \bibinfo{person}{Scott Rickard}.} \bibinfo{year}{2009}\natexlab{}.
\newblock \showarticletitle{Comparing measures of sparsity}.
\newblock \bibinfo{journal}{\emph{IEEE Transactions on Information Theory}} \bibinfo{volume}{55}, \bibinfo{number}{10} (\bibinfo{year}{2009}), \bibinfo{pages}{4723--4741}.
\newblock


\bibitem[Kang and McAuley(2018)]%
        {kang2018self}
\bibfield{author}{\bibinfo{person}{Wang-Cheng Kang} {and} \bibinfo{person}{Julian McAuley}.} \bibinfo{year}{2018}\natexlab{}.
\newblock \showarticletitle{Self-attentive sequential recommendation}. In \bibinfo{booktitle}{\emph{2018 IEEE international conference on data mining (ICDM)}}. IEEE, \bibinfo{pages}{197--206}.
\newblock


\bibitem[Klenitskiy and Vasilev(2023)]%
        {klenitskiy2023turning}
\bibfield{author}{\bibinfo{person}{Anton Klenitskiy} {and} \bibinfo{person}{Alexey Vasilev}.} \bibinfo{year}{2023}\natexlab{}.
\newblock \showarticletitle{Turning Dross Into Gold Loss: is BERT4Rec really better than SASRec?}. In \bibinfo{booktitle}{\emph{Proceedings of the 17th ACM Conference on Recommender Systems}}. \bibinfo{pages}{1120--1125}.
\newblock


\bibitem[Li et~al\mbox{.}(2021b)]%
        {li2021intention}
\bibfield{author}{\bibinfo{person}{Haoyang Li}, \bibinfo{person}{Xin Wang}, \bibinfo{person}{Ziwei Zhang}, \bibinfo{person}{Jianxin Ma}, \bibinfo{person}{Peng Cui}, {and} \bibinfo{person}{Wenwu Zhu}.} \bibinfo{year}{2021}\natexlab{b}.
\newblock \showarticletitle{Intention-aware sequential recommendation with structured intent transition}.
\newblock \bibinfo{journal}{\emph{IEEE Transactions on Knowledge and Data Engineering}} \bibinfo{volume}{34}, \bibinfo{number}{11} (\bibinfo{year}{2021}), \bibinfo{pages}{5403--5414}.
\newblock


\bibitem[Li et~al\mbox{.}(2021a)]%
        {li2021lightweight}
\bibfield{author}{\bibinfo{person}{Yang Li}, \bibinfo{person}{Tong Chen}, \bibinfo{person}{Peng-Fei Zhang}, {and} \bibinfo{person}{Hongzhi Yin}.} \bibinfo{year}{2021}\natexlab{a}.
\newblock \showarticletitle{Lightweight self-attentive sequential recommendation}. In \bibinfo{booktitle}{\emph{Proceedings of the 30th ACM International Conference on Information \& Knowledge Management}}. \bibinfo{pages}{967--977}.
\newblock


\bibitem[Liu et~al\mbox{.}(2021b)]%
        {liu2021noninvasive}
\bibfield{author}{\bibinfo{person}{Chang Liu}, \bibinfo{person}{Xiaoguang Li}, \bibinfo{person}{Guohao Cai}, \bibinfo{person}{Zhenhua Dong}, \bibinfo{person}{Hong Zhu}, {and} \bibinfo{person}{Lifeng Shang}.} \bibinfo{year}{2021}\natexlab{b}.
\newblock \showarticletitle{Noninvasive self-attention for side information fusion in sequential recommendation}. In \bibinfo{booktitle}{\emph{Proceedings of the AAAI conference on artificial intelligence}}, Vol.~\bibinfo{volume}{35}. \bibinfo{pages}{4249--4256}.
\newblock


\bibitem[Liu et~al\mbox{.}(2021a)]%
        {liu2021augmenting}
\bibfield{author}{\bibinfo{person}{Zhiwei Liu}, \bibinfo{person}{Ziwei Fan}, \bibinfo{person}{Yu Wang}, {and} \bibinfo{person}{Philip~S Yu}.} \bibinfo{year}{2021}\natexlab{a}.
\newblock \showarticletitle{Augmenting sequential recommendation with pseudo-prior items via reversely pre-training transformer}. In \bibinfo{booktitle}{\emph{Proceedings of the 44th international ACM SIGIR conference on Research and development in information retrieval}}. \bibinfo{pages}{1608--1612}.
\newblock


\bibitem[Ma et~al\mbox{.}(2020)]%
        {ma2020disentangled}
\bibfield{author}{\bibinfo{person}{Jianxin Ma}, \bibinfo{person}{Chang Zhou}, \bibinfo{person}{Hongxia Yang}, \bibinfo{person}{Peng Cui}, \bibinfo{person}{Xin Wang}, {and} \bibinfo{person}{Wenwu Zhu}.} \bibinfo{year}{2020}\natexlab{}.
\newblock \showarticletitle{Disentangled self-supervision in sequential recommenders}. In \bibinfo{booktitle}{\emph{Proceedings of the 26th ACM SIGKDD International Conference on Knowledge Discovery \& Data Mining}}. \bibinfo{pages}{483--491}.
\newblock


\bibitem[Petrov and Macdonald(2022)]%
        {petrov2022systematic}
\bibfield{author}{\bibinfo{person}{Aleksandr Petrov} {and} \bibinfo{person}{Craig Macdonald}.} \bibinfo{year}{2022}\natexlab{}.
\newblock \showarticletitle{A systematic review and replicability study of bert4rec for sequential recommendation}. In \bibinfo{booktitle}{\emph{Proceedings of the 16th ACM Conference on Recommender Systems}}. \bibinfo{pages}{436--447}.
\newblock


\bibitem[Petrov and Makarov(2021)]%
        {petrov2021attention}
\bibfield{author}{\bibinfo{person}{Aleksandr Petrov} {and} \bibinfo{person}{Yuriy Makarov}.} \bibinfo{year}{2021}\natexlab{}.
\newblock \showarticletitle{Attention-based neural re-ranking approach for next city in trip recommendations}.
\newblock \bibinfo{journal}{\emph{arXiv preprint arXiv:2103.12475}} (\bibinfo{year}{2021}).
\newblock


\bibitem[Qiu et~al\mbox{.}(2022)]%
        {qiu2022contrastive}
\bibfield{author}{\bibinfo{person}{Ruihong Qiu}, \bibinfo{person}{Zi Huang}, \bibinfo{person}{Hongzhi Yin}, {and} \bibinfo{person}{Zijian Wang}.} \bibinfo{year}{2022}\natexlab{}.
\newblock \showarticletitle{Contrastive learning for representation degeneration problem in sequential recommendation}. In \bibinfo{booktitle}{\emph{Proceedings of the fifteenth ACM international conference on web search and data mining}}. \bibinfo{pages}{813--823}.
\newblock


\bibitem[Radford et~al\mbox{.}(2018)]%
        {radford2018improving}
\bibfield{author}{\bibinfo{person}{Alec Radford}, \bibinfo{person}{Karthik Narasimhan}, \bibinfo{person}{Tim Salimans}, \bibinfo{person}{Ilya Sutskever}, {et~al\mbox{.}}} \bibinfo{year}{2018}\natexlab{}.
\newblock \showarticletitle{Improving language understanding by generative pre-training}.
\newblock  (\bibinfo{year}{2018}).
\newblock


\bibitem[Rendle et~al\mbox{.}(2010)]%
        {rendle2010factorizing}
\bibfield{author}{\bibinfo{person}{Steffen Rendle}, \bibinfo{person}{Christoph Freudenthaler}, {and} \bibinfo{person}{Lars Schmidt-Thieme}.} \bibinfo{year}{2010}\natexlab{}.
\newblock \showarticletitle{Factorizing personalized markov chains for next-basket recommendation}. In \bibinfo{booktitle}{\emph{Proceedings of the 19th international conference on World wide web}}. \bibinfo{pages}{811--820}.
\newblock


\bibitem[Singh(2020)]%
        {singh2020scalability}
\bibfield{author}{\bibinfo{person}{Monika Singh}.} \bibinfo{year}{2020}\natexlab{}.
\newblock \showarticletitle{Scalability and sparsity issues in recommender datasets: a survey}.
\newblock \bibinfo{journal}{\emph{Knowledge and Information Systems}} \bibinfo{volume}{62}, \bibinfo{number}{1} (\bibinfo{year}{2020}), \bibinfo{pages}{1--43}.
\newblock


\bibitem[Su(2023)]%
        {kexuefm-9595}
\bibfield{author}{\bibinfo{person}{Jianlin Su}.} \bibinfo{year}{2023}\natexlab{}.
\newblock \bibinfo{booktitle}{\emph{How to measure data sparsity}}.
\newblock
\urldef\tempurl%
\url{https://spaces.ac.cn/archives/9595}
\showURL{%
\tempurl}


\bibitem[Sun et~al\mbox{.}(2019)]%
        {sun2019bert4rec}
\bibfield{author}{\bibinfo{person}{Fei Sun}, \bibinfo{person}{Jun Liu}, \bibinfo{person}{Jian Wu}, \bibinfo{person}{Changhua Pei}, \bibinfo{person}{Xiao Lin}, \bibinfo{person}{Wenwu Ou}, {and} \bibinfo{person}{Peng Jiang}.} \bibinfo{year}{2019}\natexlab{}.
\newblock \showarticletitle{BERT4Rec: Sequential recommendation with bidirectional encoder representations from transformer}. In \bibinfo{booktitle}{\emph{Proceedings of the 28th ACM international conference on information and knowledge management}}. \bibinfo{pages}{1441--1450}.
\newblock


\bibitem[Sun et~al\mbox{.}(2022)]%
        {sun2022sequential}
\bibfield{author}{\bibinfo{person}{Zhongchuan Sun}, \bibinfo{person}{Bin Wu}, \bibinfo{person}{Youwei Wang}, {and} \bibinfo{person}{Yangdong Ye}.} \bibinfo{year}{2022}\natexlab{}.
\newblock \showarticletitle{Sequential graph collaborative filtering}.
\newblock \bibinfo{journal}{\emph{Information Sciences}}  \bibinfo{volume}{592} (\bibinfo{year}{2022}), \bibinfo{pages}{244--260}.
\newblock


\bibitem[Tang and Wang(2018)]%
        {tang2018personalized}
\bibfield{author}{\bibinfo{person}{Jiaxi Tang} {and} \bibinfo{person}{Ke Wang}.} \bibinfo{year}{2018}\natexlab{}.
\newblock \showarticletitle{Personalized top-n sequential recommendation via convolutional sequence embedding}. In \bibinfo{booktitle}{\emph{Proceedings of the eleventh ACM international conference on web search and data mining}}. \bibinfo{pages}{565--573}.
\newblock


\bibitem[Tong et~al\mbox{.}(2021)]%
        {tong2021pattern}
\bibfield{author}{\bibinfo{person}{Xiaohai Tong}, \bibinfo{person}{Pengfei Wang}, \bibinfo{person}{Chenliang Li}, \bibinfo{person}{Long Xia}, {and} \bibinfo{person}{Shaozhang Niu}.} \bibinfo{year}{2021}\natexlab{}.
\newblock \showarticletitle{Pattern-enhanced Contrastive Policy Learning Network for Sequential Recommendation.}. In \bibinfo{booktitle}{\emph{IJCAI}}. \bibinfo{pages}{1593--1599}.
\newblock


\bibitem[Touvron et~al\mbox{.}(2023)]%
        {touvron2023llama}
\bibfield{author}{\bibinfo{person}{Hugo Touvron}, \bibinfo{person}{Thibaut Lavril}, \bibinfo{person}{Gautier Izacard}, \bibinfo{person}{Xavier Martinet}, \bibinfo{person}{Marie-Anne Lachaux}, \bibinfo{person}{Timothée Lacroix}, \bibinfo{person}{Baptiste Rozière}, \bibinfo{person}{Naman Goyal}, \bibinfo{person}{Eric Hambro}, \bibinfo{person}{Faisal Azhar}, \bibinfo{person}{Aurelien Rodriguez}, \bibinfo{person}{Armand Joulin}, \bibinfo{person}{Edouard Grave}, {and} \bibinfo{person}{Guillaume Lample}.} \bibinfo{year}{2023}\natexlab{}.
\newblock \bibinfo{title}{LLaMA: Open and Efficient Foundation Language Models}.
\newblock
\newblock
\showeprint[arxiv]{2302.13971}~[cs.CL]


\bibitem[Vaswani et~al\mbox{.}(2017)]%
        {vaswani2017attention}
\bibfield{author}{\bibinfo{person}{Ashish Vaswani}, \bibinfo{person}{Noam Shazeer}, \bibinfo{person}{Niki Parmar}, \bibinfo{person}{Jakob Uszkoreit}, \bibinfo{person}{Llion Jones}, \bibinfo{person}{Aidan~N Gomez}, \bibinfo{person}{{\L}ukasz Kaiser}, {and} \bibinfo{person}{Illia Polosukhin}.} \bibinfo{year}{2017}\natexlab{}.
\newblock \showarticletitle{Attention is all you need}.
\newblock \bibinfo{journal}{\emph{Advances in neural information processing systems}}  \bibinfo{volume}{30} (\bibinfo{year}{2017}).
\newblock


\bibitem[Wang et~al\mbox{.}(2010)]%
        {wang2010information}
\bibfield{author}{\bibinfo{person}{Wei Wang}, \bibinfo{person}{Martin~J Wainwright}, {and} \bibinfo{person}{Kannan Ramchandran}.} \bibinfo{year}{2010}\natexlab{}.
\newblock \showarticletitle{Information-theoretic limits on sparse signal recovery: Dense versus sparse measurement matrices}.
\newblock \bibinfo{journal}{\emph{IEEE Transactions on Information Theory}} \bibinfo{volume}{56}, \bibinfo{number}{6} (\bibinfo{year}{2010}), \bibinfo{pages}{2967--2979}.
\newblock


\bibitem[Wu et~al\mbox{.}(2020)]%
        {wu2020sse}
\bibfield{author}{\bibinfo{person}{Liwei Wu}, \bibinfo{person}{Shuqing Li}, \bibinfo{person}{Cho-Jui Hsieh}, {and} \bibinfo{person}{James Sharpnack}.} \bibinfo{year}{2020}\natexlab{}.
\newblock \showarticletitle{SSE-PT: Sequential recommendation via personalized transformer}. In \bibinfo{booktitle}{\emph{Proceedings of the 14th ACM Conference on Recommender Systems}}. \bibinfo{pages}{328--337}.
\newblock


\bibitem[Wu et~al\mbox{.}(2021)]%
        {wu2021seq2bubbles}
\bibfield{author}{\bibinfo{person}{Qitian Wu}, \bibinfo{person}{Chenxiao Yang}, \bibinfo{person}{Shuodian Yu}, \bibinfo{person}{Xiaofeng Gao}, {and} \bibinfo{person}{Guihai Chen}.} \bibinfo{year}{2021}\natexlab{}.
\newblock \showarticletitle{Seq2bubbles: Region-based embedding learning for user behaviors in sequential recommenders}. In \bibinfo{booktitle}{\emph{Proceedings of the 30th ACM International Conference on Information \& Knowledge Management}}. \bibinfo{pages}{2160--2169}.
\newblock


\bibitem[Yue et~al\mbox{.}(2021)]%
        {yue2021black}
\bibfield{author}{\bibinfo{person}{Zhenrui Yue}, \bibinfo{person}{Zhankui He}, \bibinfo{person}{Huimin Zeng}, {and} \bibinfo{person}{Julian McAuley}.} \bibinfo{year}{2021}\natexlab{}.
\newblock \showarticletitle{Black-box attacks on sequential recommenders via data-free model extraction}. In \bibinfo{booktitle}{\emph{Proceedings of the 15th ACM Conference on Recommender Systems}}. \bibinfo{pages}{44--54}.
\newblock


\bibitem[Zhang et~al\mbox{.}(2020)]%
        {zhang2020match4rec}
\bibfield{author}{\bibinfo{person}{Lingxiao Zhang}, \bibinfo{person}{Jiangpeng Yan}, \bibinfo{person}{Yujiu Yang}, {and} \bibinfo{person}{Li Xiu}.} \bibinfo{year}{2020}\natexlab{}.
\newblock \showarticletitle{Match4rec: A novel recommendation algorithm based on bidirectional encoder representation with the matching task}. In \bibinfo{booktitle}{\emph{Neural Information Processing: 27th International Conference, ICONIP 2020, Bangkok, Thailand, November 23--27, 2020, Proceedings, Part III 27}}. Springer, \bibinfo{pages}{491--503}.
\newblock


\bibitem[Zhang et~al\mbox{.}(2021)]%
        {zhang2021behavioral}
\bibfield{author}{\bibinfo{person}{Yixin Zhang}, \bibinfo{person}{Lizhen Cui}, \bibinfo{person}{Wei He}, \bibinfo{person}{Xudong Lu}, {and} \bibinfo{person}{Shipeng Wang}.} \bibinfo{year}{2021}\natexlab{}.
\newblock \showarticletitle{Behavioral data assists decisions: exploring the mental representation of digital-self}.
\newblock \bibinfo{journal}{\emph{International Journal of Crowd Science}} \bibinfo{volume}{5}, \bibinfo{number}{2} (\bibinfo{year}{2021}), \bibinfo{pages}{185--203}.
\newblock


\bibitem[Zhou et~al\mbox{.}(2021)]%
        {zhou2021contrastive}
\bibfield{author}{\bibinfo{person}{Chang Zhou}, \bibinfo{person}{Jianxin Ma}, \bibinfo{person}{Jianwei Zhang}, \bibinfo{person}{Jingren Zhou}, {and} \bibinfo{person}{Hongxia Yang}.} \bibinfo{year}{2021}\natexlab{}.
\newblock \showarticletitle{Contrastive learning for debiased candidate generation in large-scale recommender systems}. In \bibinfo{booktitle}{\emph{Proceedings of the 27th ACM SIGKDD Conference on Knowledge Discovery \& Data Mining}}. \bibinfo{pages}{3985--3995}.
\newblock


\bibitem[Zhou et~al\mbox{.}(2020)]%
        {zhou2020s3}
\bibfield{author}{\bibinfo{person}{Kun Zhou}, \bibinfo{person}{Hui Wang}, \bibinfo{person}{Wayne~Xin Zhao}, \bibinfo{person}{Yutao Zhu}, \bibinfo{person}{Sirui Wang}, \bibinfo{person}{Fuzheng Zhang}, \bibinfo{person}{Zhongyuan Wang}, {and} \bibinfo{person}{Ji-Rong Wen}.} \bibinfo{year}{2020}\natexlab{}.
\newblock \showarticletitle{S3-rec: Self-supervised learning for sequential recommendation with mutual information maximization}. In \bibinfo{booktitle}{\emph{Proceedings of the 29th ACM international conference on information \& knowledge management}}. \bibinfo{pages}{1893--1902}.
\newblock


\bibitem[Zhou et~al\mbox{.}(2022)]%
        {zhou2022filter}
\bibfield{author}{\bibinfo{person}{Kun Zhou}, \bibinfo{person}{Hui Yu}, \bibinfo{person}{Wayne~Xin Zhao}, {and} \bibinfo{person}{Ji-Rong Wen}.} \bibinfo{year}{2022}\natexlab{}.
\newblock \showarticletitle{Filter-enhanced MLP is all you need for sequential recommendation}. In \bibinfo{booktitle}{\emph{Proceedings of the ACM web conference 2022}}. \bibinfo{pages}{2388--2399}.
\newblock


\end{thebibliography}
